\documentclass[prd,twocolumn,superscriptaddress,showpacs,floatfix,aps]{revtex4-2}
\usepackage[english]{babel}
\usepackage[utf8]{inputenc}
\usepackage{graphicx,amsfonts,amssymb,amsmath,xspace,physics}
\usepackage{mathtools}
\usepackage{comment}
\usepackage{enumitem}
\usepackage[pdftex,breaklinks=true,colorlinks=true]{hyperref}
\usepackage{epstopdf}
\usepackage{tabularx}
\usepackage{array}
\usepackage{multirow}
\usepackage{diagbox}
\usepackage{enumitem}
\usepackage[dvipsnames]{xcolor,colortbl}
\usepackage{stmaryrd}
\usepackage{tikz}
\usepackage{ulem}
\usepackage{soul}
\usepackage{orcidlink}
\usepackage{float}

\usetikzlibrary{backgrounds}
\pgfdeclarelayer{background}
\pgfsetlayers{background,main}

\usetikzlibrary{tikzmark,calc,patterns}
\definecolor{g}{rgb}{.1,0.4,.1}
\definecolor{b}{rgb}{0,0.2,1}
\definecolor{rouge}{rgb}{0.82,0.,0.}
\definecolor{vert}{rgb}{0.,0.82,0.}
\definecolor{darkgreen}{rgb}{0.,0.75,0.}

\definecolor{orange}{rgb}{1,0.5,0.}
\definecolor{bleu}{rgb}{0.,0.,0.82}
\definecolor{m}{rgb}{0.82,0.,0.82}
\definecolor{vert2}{rgb}{0.,0.5,0.}
\definecolor{rougeclair}{rgb}{1.0,0.7,0.7}
\definecolor{Gray}{gray}{0.85}

\newcommand{\xdownarrowright}[2][]{%
  \mathrel{\begin{tikzpicture}[baseline=-0.4ex]
    \draw[->] (0,0) -- (0,-#2);
    \node[right] at (0,-0.5*#2) {\scriptsize $#1$};
  \end{tikzpicture}}%
}

\newcommand{\beq}{\begin{equation}}
\newcommand{\be}{\begin{equation}}
\newcommand{\beqn}{\begin{eqnarray}}
\newcommand{\eeq}{\end{equation}}
\newcommand{\ee}{\end{equation}}
\newcommand{\eeqn}{\end{eqnarray}}

\newcommand{\bem}{\begin{pmatrix}}
\newcommand{\eem}{\end{pmatrix}}

\newlength{\ldag}
\newcommand{\older}[1]{{\leavevmode\color{blue}}}

\begin{document}

\title{Noise tailoring for error mitigation and for diagnozing digital quantum computers}

\author{Thibault Scoquart\normalfont\,\orcidlink{0000-0002-0449-6344}\textsuperscript{\fnsymbol{footnote}{$\dagger$}}}
\email{thibault.scoquart@irsamc.ups-tlse.fr}
\affiliation{Institute for Quantum Materials and Technologies, Karlsruhe Institute of Technology, 76131 Karlsruhe, Germany}
\affiliation{Laboratoire de Physique Théorique, Université de Toulouse, CNRS, France}

\author{Hugo Perrin\normalfont\,\orcidlink{0000-0002-1313-8549}\textsuperscript{\fnsymbol{footnote}{$\dagger$}}}
\affiliation{Institute for Quantum Materials and Technologies, Karlsruhe Institute of Technology, 76131 Karlsruhe, Germany}
\affiliation{University of Strasbourg and CNRS, CESQ and ISIS (UMR 7006), aQCess, Strasbourg, France}
\affiliation{QPerfect SAS, 67200, Strasbourg, France}

\author{Kyrylo Snizhko\normalfont\,\orcidlink{0000-0002-7236-6779}}
\affiliation{Univ. Grenoble Alpes, CEA, Grenoble INP, IRIG, PHELIQS, 38000 Grenoble, France}

\date{\today}

\begin{abstract}
Error mitigation (EM) methods are crucial for obtaining reliable results in the realm of noisy intermediate-scale quantum (NISQ) computers, where noise significantly impacts output accuracy. Some EM protocols are particularly efficient for specific types of noise. Yet the noise in the actual hardware may not align with that. In this article, we introduce Noise Tailoring (NT)---an innovative strategy designed to modify the structure of the noise associated with two-qubit gates through statistical sampling. We perform classical emulation of the protocol behavior and find that the NT+EM results can be up to 5 times more accurate than the results of EM alone for realistic Pauli noise acting on two-qubit gates. At the same time, on actual IBM quantum computers, the NT method falls victim to various small error sources beyond Markovian Pauli noise. We propose to use the NT method for characterizing such error sources on quantum computers in order to inform hardware development.
\end{abstract}

\maketitle

\begingroup
\renewcommand{\thefootnote}{\fnsymbol{footnote}} 
\setcounter{footnote}{0} 
\insert\footins{%
  \footnotesize
  \interlinepenalty=10000
  \hsize=\columnwidth
  \parindent=1em
  \noindent\textsuperscript{\fnsymbol{footnote}{$\dagger$}} These authors contributed equally to this work.\par
}
\endgroup

\section{Introduction}
\par The era of Noisy Intermediate-Scale Quantum (NISQ) devices~\cite{Preskill2018} marks a crucial phase in quantum computing. NISQ devices, characterized by their imperfect quantum operations, present a unique blend of opportunities and challenges. Their emergence has sparked a surge in research aimed at harnessing their potential. The imperfections inherent in NISQ technology, such as qubit decoherence and operational errors, are major hurdles, but they also drive the quest for innovative solutions in quantum computing.

One of the most promising avenues to utilize the capabilities of NISQ platforms lies in error mitigation (EM) protocols~\cite{Endo2021,Cai2023}. Further, EM protocols can have utility, when used with early fault-tolerant quantum computers~\cite{Aharonov2025}. These protocols are designed to reduce the impact of errors in quantum computations without the need for full-scale quantum error correction, which is beyond current capabilities. Among the various strategies, Probabilistic Error Cancellation (PEC)~\cite{vandenberg2023,temme2017} and Zero Noise Extrapolation (ZNE)~\cite{temme2017,dumitrescu2018,giurgica-tiron2020,he2020} have emerged as the frontrunners. PEC operates by inverting the effects of noise, while ZNE infers the desired noise-free result from multiple results at various levels of noise strength via extrapolation. Each method comes with its trade-offs; PEC requires detailed noise knowledge and is resource intensive~\cite{Blume-Kohout2013,Blume-Kohout2017,Endo2018,Zhang2020,Nielsen2021}, whereas ZNE, though less resource-demanding, does not fully tackle all noise, leading to biased estimates of the noiseless expectation value \cite{Endo2018,Kurita2023,Pelofske2024, Liao2025}.

Alternatively, other error mitigation techniques (EM) exist. They are often efficient for particular types of noise structure. For example, the Noise Estimation Circuit (NEC) method~\cite{urbanek2021} is efficient in handling isotropic, depolarizing noise. Yet the actual hardware noise may have a different structure.

The integration of different error mitigation techniques appears promising for boosting their efficacy~\cite{McArdle2019,bultrini2023}. Combining PEC and ZNE leads to various new error mitigation strategies known as probabilistic error reduction (PER)~\cite{mari2021,mcdonough2022}, probabilistic error amplification~\cite{Li2017,mari2021,kim2023} or noise-extended probabilistic error cancellation (NEPEC)~\cite{mari2021} methods. While the full noise cancellation via PEC encounters scalability challenges with increasing noise strength and/or circuit size, PER enables choosing the desired noise strength, while controlling the sampling overhead. ZNE is then employed to infer the true result from several imperfect PER results at various noise levels. This idea can be further extended by combining Probabilistic Error Amplification (PEA) in combination with ZNE, further reducing the circuit sampling cost, and producing promising levels of accuracy \cite{kim2023}.

This article seeks to extend and generalize these approaches to other EM methods, beyond ZNE. Specifically, the EM methods that assume specific simple noise channels. We introduce a novel strategy termed Noise Tailoring (NT). NT is a sampling technique similar to PER, yet whose aim is to bring the noise to a desired \textit{structure}, cf.~Fig.~\ref{fig:workflow}, while the noise strength can be chosen to minimize the sampling cost. The modified noise shape can be chosen to maximize the efficiency of the specific EM method used.

We illustrate our approach by coupling NT with NEC. We apply the NT+NEC method for simulating a quench in the BCS model. We compare this protocol to our previous work that used only NEC~\cite{perrin2024}. In classical simulations, we show the potential for improvement due to NT by a factor $\sim 5$. When performing experiments on actual NISQ devices, we only achieve worse results with NT+NEC protocol in comparison to pure NEC. We attribute the discrepancy to additional sources of noise that are present on the NISQ devices, but are not accounted for in our classical simulations. At the same time, an in-depth analysis of the NISQ data shows that increasing the number of sampling circuits in the NT protocol would allow one to improve the accuracy of the results by a factor $\sim 2$ compared to the protocol not using NT.

Most importantly, the same in-depth analysis of the NISQ data enables gaining qualitative and quantitative insights about of these additional sources of noise. Such errors are inherently small, yet are amplified by the application of the NT protocol (they would also be amplified by PER or PEC). We propose using our protocol for characterizing such small errors, the information about which could then be used to inform hardware development.

The paper is organized as follows. We explain the NT method and the whole stack of techniques we use in Sec.~\ref{sec:NT_to_depo}. In Sec.~\ref{sec:classical_sims}, we demonstrate the expected benefits of the NT technique based on classical  emulation of quantum computer performance. We present the results of actual NISQ experiments on IBM quantum computers in Sec.~\ref{sec:NISQ}. We discuss the potential value of the NT technique in various contexts in Sec.~\ref{sec:Discussion}. We provide a brief conclusion in Sec.~\ref{sec:conclusion}.

\section{Noise-tailoring for a 2-qubit noise channel}
\label{sec:NT_to_depo}

\begin{figure*}
    \centering
    \includegraphics[scale=0.8]{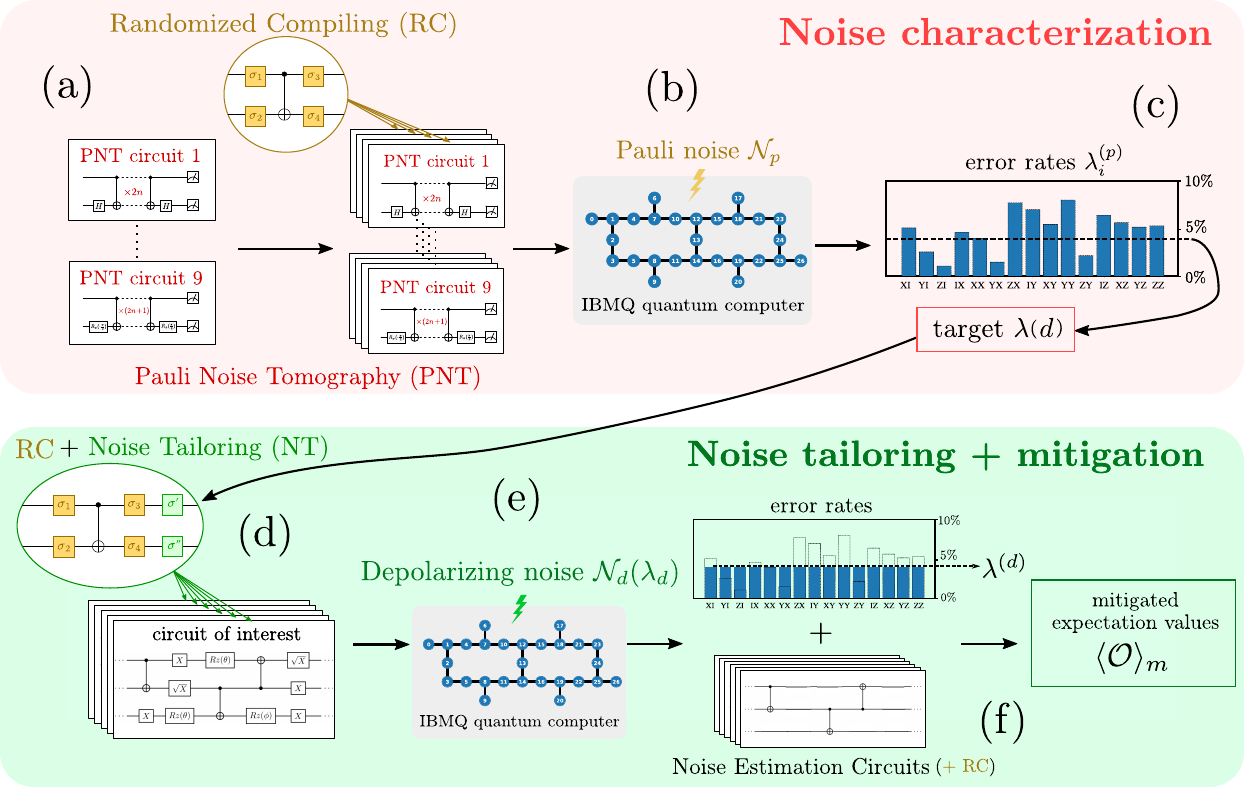}
    \caption{Workflow diagram describing the implementation of Noise Tailoring (NT) and how we combine it with other techniques in our protocol. (a) Randomized Compiling (RC) is performed in order to convert the error channel of CNOT gates to Pauli error channel. Pauli Noise Tomography (PNT) is then performed in order to characterize this Pauli error channel. (b) Due to RC, the average outputs of the circuits correspond to an effective Pauli noise acting on CNOT gates. This is the noise assumed by the rest of the protocol. (c) The knowledge of the Pauli noise coefficients is used to determine the target noise channel to be aimed for by NT. (Here for illustration, the target channel is the depolarizing noise whose strength is the average of all the Pauli coefficients). (d) RC and NT are applied simultaneously to the circuit of interest. The probability of applying various single-qubit Pauli gates for the NT protocol is computed based on the Pauli noise characterized in the previous steps and the target noise chosen. (e) As a result of the RC+NT combination, the effective noise acting on the CNOT gates of the circuit is the target noise (which is chosen to be depolarizing here). (f) The obesrvables of interest are measured. The efficiency of the error mitigation scheme (here, Noise Estimation Circuits --- NEC) is enhanced by the noise structure adjustment due to the use of NT. The NT technique used in the protocol constitutes the main novelty of the present work.}
    \label{fig:workflow}
\end{figure*}

Here we give a general overview of our method. Our protocol combines Randomized compiling (RC), Pauli Noise Tomography (PNT), the NT technique (which is the key novelty of the present work), and NEC. The combination of RC and NT effectively converts experimental noise on 2-qubit gates to a target noise channel, which we choose for optimizing the performance of the NEC technique. The workflow of our method is presented schematically in Fig.~\ref{fig:workflow}. We explain this workflow in detail in this section.

For the sake of simplicity, we discuss here the basic version of our protocol, which discards spillover crosstalk and only concerns the active qubits of the CNOT gates. In fact, we have generalized the protocol to account for nearest-neighbor crosstalk effects; we present it in Appendix~\ref{app:crosstalk}. It is this generalized version of our protocol that we utilize in Secs.~\ref{sec:classical_sims} and \ref{sec:NISQ}. We denote these crosstalk-aware parts of our protocol as cRC and cNT.

\subsection{Raw noise on the quantum computer}
\label{sec:raw_noise}

On IBMQ quantum computers~\cite{ibmquantum}, and more generally on NISQ, the dominant source of errors is the imperfect application of 2-qubit entangling gates~\cite{kandala2021}. State preparation and measurement (SPAM) errors have a magnitude of a few percent, but they only manifest once per circuit. When working with deep circuits, this makes the 2-qubit gate errors dominant by a huge margin. In the case of IBMQ superconducting quantum computers, the basis gate for 2-qubit operations is the CNOT gate, whose error rate, estimated using cycle benchmarking \cite{Erhard2019}, is of the order of $1\%$ \footnote{At the time where we performed the experiments presented here, IBMQ devices could implement a single type of native 2-qubit gate, the CNOT gate, with an estimated error rate of the order of $1\%$. Since then, they have been replaced by the CZ gate, which is equivalent to a CNOT gate up to single-qubit rotations, and allow for slightly lower error rates, of the order of $0.4\%$ on average.}. The effect of these experimental imperfections can be modeled by a 2-qubit quantum noise channel---different channel for each junction between pairs of physical qubits---which is a mixture of coherent (systematic errors) and incoherent (stochastic errors) noise channels. In this work, we denote this noise channel by $\mathcal{N}_0$.

Our protocol is designed under two main assumptions:
\begin{enumerate}[label=(\roman*)]
    \item The noise is Markovian, such that when running a quantum circuit with multiple CNOT gates, the probabilities of errors occuring on each CNOT are completely independent.
    \item Single qubit gates are implemented perfectly on the quantum device. On IBMQ quantum computers (QCs), their measured error rate is one to two orders of magnitude below the CNOT gates' error rate.
\end{enumerate}
These assumptions are only met approximately on a real quantum device, which contributes to the infidelity associated with a real quantum computation. This contribution becomes apparent when one compares classical emulations (Sec.~\ref{sec:classical_sims}), in which both (i) and (ii) are exact, to quantum results (Sec.~\ref{sec:NISQ}), all other things being equal.

In addition to being very complex, experimental noise can vary significantly in strength and structure over time scales of the order of a few hours \cite{woitzik2024,Dasgupta2024,Filippov2024}, making it difficult to measure the noise characteristics and use them for NT. Indeed, any information about the noise channels acting on CNOT gates needed to implement error mitigation schemes must be gathered right before each experiment, in a timely manner.

\subsection{From raw noise to effective depolarizing noise}

We are now in position to describe the sequence of techniques that constitute our protocol, which turns the raw noise channel $\mathcal{N}_0$ acting on the CNOT gates of a NISQ device into a purely depolarizing noise channel. Under our assumptions (i) and (ii), this mapping is exact in the limit of infinite circuit sampling, which we consider here. The global circuit overhead and associated deviations from the infinite-sampling case will be discussed in Sec.~\ref{sec:protocol_on_NISQ}.

\subsubsection{Randomized Compiling (RC)}

 We start by considering a single junction between two superconducting qubits, on which CNOT gates may be applied during the execution of a quantum circuit of interest. As a first step for our method, we remove the coherent part of $\mathcal{N}_0$ using Randomized Compiling (RC) \cite{wallman2016,kern2005,hashim2021,cai2019}, a standard method for NISQ. RC is a sampling technique that involves running additional versions of a quantum circuit of interest, in which all CNOT gates have been dressed with random combinations of single-qubit Pauli gates, with the constraint that the dressed versions of the CNOT gate still implement a logical CNOT gate operation, cf.~Fig.~\ref{fig:workflow}(a,d).

 It can be shown that averaging the outputs (e.g., the expectation value of a measured local observable) over multiple circuits with randomly dressed CNOT gates yields the effective output that would have been obtained from the original circuit under a purely stochastic 2-qubit Pauli noise channel. For each junction, the noise channel is thus transformed as:
\begin{align}
\mathcal{N}_0 \xrightarrow{\,\,\text{RC}\,\,} \mathcal{N}_p^\infty(\rho)=\sum_{i=0}^{15} \lambda^{(p)}_{i}\sigma_i\,\rho\,\sigma_i,
\label{eq:pauli_noise_RC_infinite_sampling}
\end{align}
where $\sigma_i\in\{\mathbb{I},X, Y, Z\}^{\otimes 2}$ ($X$, $Y$ and $Z$ being Pauli matrices), and $\sum_{i}\lambda^{(p)}_{i}=1$. Here and throughout this work, the superscript $\infty$ refers to the effective result of an infinite circuit sampling for a given procedure. The effective error rates $\lambda^{(p)}_{i}$ correspond to the incoherent part of the noise obtained after the contribution of coherent noise (involving terms of the form $\sigma_i\,\rho\,\sigma_j$, $i\neq j$) has been exactly suppressed.

For the sake of discussion, let us now assume that one wants to perform a quantum simulation of a given circuit of interest, which contains $N_\text{CNOT}$ noisy CNOT gates, distributed across $N_q$ qubits (or equivalently $N_q-1$ junctions, for a linear layout of qubits). The goal is to measure the expectation value of a generic observable $\mathcal{O}$ with the best possible accuracy. Averaging over a large number of sampling circuits $N_s = \infty$, the circuit output transforms the expectation value of a generic observable $\hat{O}$:

\begin{align}
  \langle \hat{O} \rangle_0 = \text{Tr}(\rho_0 \hat{O} ) \xrightarrow{\,\,\text{RC}\,\,}  \langle \hat{O} \rangle_p^\infty =  \text{Tr}(\rho_p^\infty \hat{O} ),
\label{eq:exp_value_RC_infinite_sampling}
\end{align}
where $\rho_0$ is the output density matrix of the circuit with raw noise ($\mathcal{N}_0$) on the CNOT gates and $\rho_p^\infty$ is the output density matrix of the circuit with the Pauli noise $\mathcal{N}_p^\infty$ on the CNOT gates. This is the effect of averaging over all possible RC circuits.

\subsubsection{Pauli Noise Tomography (PNT)}
\label{subsubsec:PNT}

Provided that every circuit we run is Randomly Compiled, we are assured that on each junction, the effective noise channel acting on CNOT gates has a Pauli noise structure $\mathcal{N}_p^\infty$. In order to use the next steps in our protocol, we require the knowledge of the Pauli noise error rates $\{\lambda^{(p)}_i\}_{i\in [1,15]}$. These can be obtained using Pauli Noise Tomography (PNT) on each qubit junction
\cite{mcdonough2022,vandenberg2023,Chen2023},
cf.~Fig.~\ref{fig:workflow}(a-c). For the reader's convenience, we describe the PNT procedure explicitly in Appendix~\ref{app:PNT}.

We note that, in principle, when SPAM errors are present, Pauli noise can only be learned up to a gauge freedom, with different parameter choices yielding identical observable outcomes \cite{Chen2023, Hines2025}. Recent investigations show that self-consistent characterization methods that jointly learn gate and SPAM errors ensure that predictions and error mitigation strategies remain unbiased regardless of the specific gauge choice \cite{Chen2025}. In this work, we mitigate measurement errors by employing a readout error correction (REC) scheme based on the iterative Bayesian unfolding method ~\cite{nachman2020} and determine Pauli coefficients based on these mitigated PNT results. We acknowledge that this approach may introduce systematic biases as SPAM and Pauli errors are not characterized jointly. However, we employ deep circuits where we expect SPAM errors to be negligible relative to accumulated gate errors.

\subsubsection{Noise-Tailoring}
\label{sec:NT_infinite_sampling}

The main novelty of this work is the use of the NT technique we discuss here. NT is a generalization of PEC/PER. While PEC and PER typically aim to reduce the noise amplitude (to $0$ in the case of PEC), cf.~Fig.~\ref{fig:noise_control}, NT uses a similar sampling method, but aims to tailor the noise to an arbitrarily chosen Pauli noise channel
\begin{align}
\mathcal{N}_{\rm target}^\infty(\rho) = \left(1-\sum_{i=1}^{15}\lambda^{\rm target}_i\right)\rho +  \sum_{i=1}^{15} \lambda^{\rm target}_i \sigma_i\,\rho\,\sigma_i.
\end{align}
Here the error rates error rates $\lambda^{\rm target}_i$ do not have to be related to the original noise channel in any way. In particular, some may be larger than the original $\lambda^{(p)}_{i}$, while others may be smaller. This enables choosing $\mathcal{N}_{\rm target}^\infty$ in order to maximize the performance of a specific EM technique, as we discuss in Sec.~\ref{subsec:NT+error_mitig}.

In order to convert $\mathcal{N}_p^\infty$ to $\mathcal{N}_{\rm target}^\infty$, one should prepare alternative, random versions of our circuit of interest. In these random circuits, the CNOT gates are dressed with extra Pauli gates---on top of the RC gates, cf.~Fig.~\ref{fig:workflow}(d). These extra Pauli gates should be drawn randomly with a carefully chosen probability distribution that is calculated based on error rates of the Pauli noise channel acting on a given junction, $\{\lambda^{(p)}_i\}_{i\in [1,15]}$, and the \textit{desired} error rates $\{\lambda^{\rm target}_i\}_{i\in [1,15]}$. Averaging over those circuits yields the effective noise channel $\mathcal{N}_{\rm target}^\infty$, cf.~Fig.~\ref{fig:workflow}(d-f).

In general, the dressing Pauli operations should be drawn not from a probability distribution, but from a \textit{quasi}probability distribution $q_i$ with some $q_i < 0$; $i$ enumerates the dressing Pauli operations. In order to sample the quasiprobability distribution, one samples a probability distribution $p_i = \abs{q_i}/\gamma$ with $\gamma=\sum_i |q_{i}|>1$, then takes the signs of $q_i$ into account when averaging and multiplies the result by $\gamma$, cf.~Appendix~\ref{app:NT} for details. This peculiarity is not unique to NT---it is also present for PER and PEC, and is neatly avoided by PEA. This multiplication of the result by $\gamma > 1$ will have important consequences below.

We emphasize that the probability distribution and, correspondingly, $\gamma$ is different for each junction where CNOT gates can be applied. This is because the Pauli noise $\mathcal{N}_p^\infty$ is different on each junction, and the target noise $\mathcal{N}_{\rm target}^\infty$ may also be chosen different for different junctions.

In this work, we focus on the case where the target noise channel for each junction is a depolarizing noise channel $\mathcal{N}_d^\infty$, described by a single coefficient $\lambda^{\rm target}_i = \lambda_{d}$. In the limit of infinite circuit sampling, NT exactly maps Pauli noise to the desired Pauli noise channel (see Fig.~\ref{fig:noise_control}):
\begin{align}
\mathcal{N}_p^\infty \xrightarrow{\,\,\text{NT}\,\,} \mathcal{N}_d^\infty(\rho) = (1-15\lambda_d)\rho + \lambda_d \sum_{i=1}^{15} \sigma_i\,\rho\,\sigma_i.
\label{eq:depo_noise_infinite_sampling}
\end{align}
We keep the freedom of choosing $\lambda_d$ different for different junctions.

The output of our circuit of interest, averaged over an infinite number of alternative versions of the circuit with RC- and NT-dressed CNOT gates, yields:
\begin{align}
  \langle \hat{O} \rangle_0 \xrightarrow{\,\,\text{RC+NT}\,\,} \langle \hat{O} \rangle_d^\infty = \Tr (\rho_d^\infty \hat{O}),
\label{eq:exp_value_RC_NT_infinite_sampling}
\end{align}
where we keep using the same notation principles as in Eq.~\eqref{eq:exp_value_RC_infinite_sampling}. More details about the practical implementation of our NT technique can be found in Appendix~\ref{app:NT}.

\subsubsection{NEC}
\label{sec:NEC}

The steps presented above simplify the noise structure, yet they do not mitigate the effects of noise. In order to mitigate the errors, we use a simple error mitigation protocol---NEC \cite{urbanek2021}.

NEC consists in running, in addition to the circuit of interest, its alternative version, where all single-qubit gates have been removed. With no single-qubit gates, the circuit is a Clifford circuit that can be efficiently simulated classically for perfect CNOTs. The output from the NEC in the presence of noise, when compared to the perfect result, provides the circuit fidelity $\mathcal{F}_\text{NEC}$. Given the assumption of the dominant error stemming from CNOT gates, $\mathcal{F}_\text{NEC}$ can be used as a proxy for the fidelity of the circuit of interest (which has the same structure of CNOT gates). This can then be used to mitigate the output of the original circuit $\langle \hat{O} \rangle$ of interest as
\begin{align}
    \langle \hat{O}\rangle_m = \frac{\langle \hat{O} \rangle}{\mathcal{F}_\text{NEC}},
\end{align}
where $\langle \hat{O}\rangle_m$ is the mitigated expectation value.

NEC is thus exact for \textit{global} depolarizing noise, as the effect of such noise is purely multiplicative and identical on any circuit, including the NEC circuit and the circuit of interest. Global depolarizing noise is not realistically attainable. However, one can expect \textit{local} depolarizing noise (which can be achieved with NT) to be more suitable for employing NEC than generic Pauli noise.

Combining NT with NEC, the final mitigated observable will be obtained by multiplying the averaged output of random circuits by a factor
\begin{align}
    \sigma= \frac{\gamma^{N_\text{CNOT}}}{\mathcal{F}_\text{NEC}}.
    \label{eq:prefactor_NT_NEC}
\end{align}

Here $\gamma$ is the sampling factor associated to each CNOT gate in the NT procedure (see Sec.~\ref{sec:NT_infinite_sampling}).

\subsubsection{Choosing the optimal noise strength $\lambda_d$}
\label{sec:Choosing_optimal_target_noise}

The factor $\sigma$, introduced in Eq.~\eqref{eq:prefactor_NT_NEC}, determines the error of the mitigated result. Indeed, suppose the unmitigated expectation value $\langle \hat{O} \rangle$ has a given precision $\varepsilon$ (due to shot noise, finite sampling of probability distributions, unaccounted error sources, ...). Then the mitigated expectation value $\langle \hat{O}\rangle_m$---while more accurate due to NEC mitigating some of the noise contributions---would have precision $\sigma\varepsilon$.

Therefore, minimizing $\sigma$ is of interest for maximizing the protocol performance. This provides a guideline for choosing the target noise strength, $\lambda_d$. Indeed, both the numerator and denominator of Eq.~\eqref{eq:prefactor_NT_NEC} are expected to be decreasing functions of the depolarizing parameter $\lambda_d$, with $\lambda_d = 0$ implying $\gamma^{N_\text{CNOT}} \gg 1$, and the maximum allowed $\lambda_d = 1/15$ leading to $\mathcal{F}_\text{NEC} = 0$. Therefore, one expects an optimal value of $\lambda_d \in [0, 1/15]$ to exist.

The minimization of $\sigma$ can be performed fully classically. Indeed, $\gamma(\lambda_d)$ is classically computed from the determination of the quasiprobabilities for the NT protocol, while the noise estimation circuit, yielding $\mathcal{F}_\text{NEC}(\lambda_d)$, is a Clifford circuit that can efficiently simulated classically \cite{Takahashi2021, Nelson2024}. Thus, we choose the optimal strength of target noise $\lambda_d^\text{opt}$ for each 2-qubit junction such that $\sigma$ is minimized.

\subsection{What to expect from the protocol on a realistic NISQ device}
\label{sec:protocol_on_NISQ}

Above, we have described our protocol assuming the ideal case of infinite sampling, i.e., assuming that we can run infinitely many circuits to perform RC, PNT and NT. In practice, on NISQ devices, one can only run a large but limited number of circuits in a reasonable amount of time \footnote{The actual running time per circuit differs by orders of magnitude between various quantum computing platforms (superconducting qubits, Rydberg, cold ions etc.). For example, that what takes minutes on a superconducting platform, can take hours on Rydberg platforms.}. Therefore, the effective noise channels acting on each junction can only be approximations of the perfect infinite-sampling channels.

In this section, we discuss the overall sampling cost of each step of the method. As we do this, we introduce the notations for the finite-sampling versions of the different noise channels and the corrections to the infinite sampling limit. These notations will be useful for the discussions and result analysis in the upcoming sections. However, we do not provide formal definitions for these notations.

\subsubsection{Finite-sampling RC}

To achieve the perfect RC  of a single CNOT gate, one must average over all $16$ different dressings of the CNOT with Pauli gates that leave the CNOT operation invariant. Thus, a given quantum circuit has $N_{\mathrm{RC}}^\infty = 16^{N_\text{CNOT}}$ possible randomly compiled variants, $N_\text{CNOT}$ being the total number of CNOT gates in the quantum circuit of interest. This exponentially large number of circuits is not realistic to sample. In the limit of finite sampling, part of the coherent noise present in $\mathcal{N}_0$ is not suppressed, and the resulting noise channel acting on the CNOT gates of a given junction can be expressed as

\begin{align}
\mathcal{N}_0 \xrightarrow[N_\text{RC}\text{ circuits}]{\,\,RC\,\,} \mathcal{N}_p \equiv (1-\delta_\text{RC}\,)\mathcal{N}_p^\infty + \delta_\text{RC}\,\mathcal{N}^*_c ,
\label{eq:pauli_noise_finite_sampling}
\end{align}
where $\mathcal{N}^*_c$ is a unitary coherent noise channel with small magnitude, and $\delta_\text{RC}\sim 1/\sqrt{N_\text{RC}}$ is controlled by the number of randomly compiled circuits used, $N_{\mathrm{RC}}$.
The final expectation value obtained from the circuit of interest is also affected and yields similarly:
\begin{align}
    \langle \hat{O} \rangle_0  \xrightarrow[N_\text{RC}\text{ circuits}]{\,\,RC\,\,} \langle \hat{O} \rangle_p =\langle \hat{O} \rangle_p^\infty + \Delta\langle \hat{O} \rangle_c + \Delta\langle \hat{O} \rangle_\text{unk.},
\label{eq:expectation_value_RC_finite_sampling}
\end{align}
where $\Delta\langle \hat{O} \rangle_c \sim \delta_\text{RC}$ is the correction associated with the remaining coherent noise on all different junctions, which scales as $1/\sqrt{N_{\mathrm{RC}}}$. We also introduce $\Delta\langle \hat{O} \rangle_\text{unk.}$, which stems from unknown error sources, such as non-Markovianity of the noise and presence of single-qubit noise; the latter respectively violate our assumptions (i) and (ii) in Sec.~\ref{sec:raw_noise}.

\subsubsection{Finite-sampling PNT}

Next, we perform Pauli Noise Tomography with a finite number of circuits to gather the Pauli noise coefficients of $\mathcal{N}_p$ for each physical qubit junction on which CNOT gates will be applied. In our version of PNT, we only require $9 n_d$ carefully chosen quantum circuits to perform the full tomography of Pauli noise, $n_d\geq 2$ being a parameter that controls the precision of the tomography. We describe the PNT procedure explicitly in Appendix~\ref{app:PNT}. Of course, all PNT circuits also have to be randomly compiled.

For the PNT procedure, the estimated circuit cost is $N_{\mathrm{PNT}} = 9 n_d \times 2 N_j \times \tilde{N}_{\mathrm{RC}}$, where $N_j$ is the number of junctions, the factor $2$ refers to the two possible orientations of the CNOT gates on a given junction, and $\tilde{N}_{\mathrm{RC}}< N_{\mathrm{RC}}$ is the number of RC circuits used for the tomography. In our considerations in Secs.~\ref{sec:classical_sims} and \ref{sec:NISQ}, qubits are arranged in a linear architecture, the number of junctions used is $N_j = N_q - 1 = 2$, where $N_q =3$ is the number of qubits. In practice, using $n_d=5$, each circuit randomly compiled $\tilde{N}_{\mathrm{RC}} = 200$ times, we achieved the PNT on every junction with reasonable precision $\sigma_\text{PNT}\sim 10^{-2}$--$10^{-3}$, estimated by bootstrapping methods.

Note that the circuit overhead from this part of the protocol is negligible in comparison to the NT procedure, see below. For these reasons, in the rest of this paper we neglect the uncertainty on the evaluation of the Pauli noise parameters via PNT, and assume the exact knowledge of $\mathcal{N}_p^\infty$ for each junction.

\subsubsection{Finite-sampling NT}

Finally, we perform NT with a finite number of circuits $N_\text{NT}$, towards a  target depolarizing noise channel $\mathcal{N}_d^\infty$. As already mentioned, NT is a generalization of the well-known PEC/PER sampling procedures that targets an arbitrary effective noise channel (depolarizing channel in our case). The overall sampling cost of the procedure is controlled by a sampling factor $\gamma$, and scales approximately as $N_\text{NT} \sim \gamma^{2N_\text{CNOT}}/\epsilon^2$, $\epsilon$ being the desired accuracy on the final observable $\hat{O}$ \cite{vandenberg2023}, cf.~Secs.~\ref{sec:NT_infinite_sampling} and \ref{sec:Choosing_optimal_target_noise}. The precise value of $\gamma$ varies between different junctions of the device, as well as from one quantum device to another, depending on the actual hardware noise and the target noise.

\begin{figure}[t]
    \centering
    \includegraphics[scale=0.6]{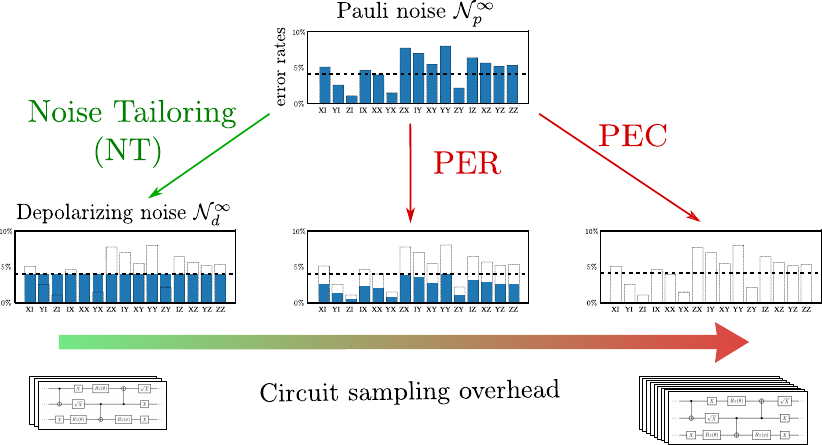}
    \caption{Illustration of different sampling approaches to control noise of 2-qubit gates. While PEC attempts to completely cancel the noise (right), PER partially reduces it, while keeping the structure (center). Both techniques typically come with a very large circuit overhead. Our novel NT approach aims to map a given Pauli noise channel to another one, selected by the user. We illustrate this in the figure by targeting the depolarizing noise channel, whose Pauli error rates are all equal (for the purpose of illustration, their value is set to the average of the original Pauli rates---dashed horizontal line on all plots). Depending on the chosen target channel, the sampling overhead might be moderate.}
    \label{fig:noise_control}
\end{figure}

This is why PEC, aiming to cancel the noise completely, naturally comes with a very large circuit overhead, as illustrated in Fig.~\ref{fig:noise_control}.

Note that since NT acts on the effective Pauli noise obtained via RC, both procedures must be implemented at once. This is illustrated in Fig.~\ref{fig:workflow}(d). We thus have $N_{NT}=N_{RC}$.

As was the case for the RC, the NT sampling is limited to a finite number of circuits, thus only approximately realizing the mapping to the target depolarizing noise channel. Consequently, a realistic implementation of NT with $N_\text{NT}$ circuits will yield a Pauli noise channel that is close to a depolarizing noise channel, which we write as $(1-\delta_\text{NT}\,)\mathcal{N}_d^\infty + \delta_\text{NT}\,\mathcal{N}^*_{p}$, where $\mathcal{N}^*_{p}$ contains all the deviations of the error rates $\lambda_i$ from the depolarizing parameter $\lambda_d$. The deviations are controlled by $\delta_\text{NT}\sim 1/\sqrt{N_\text{NT}}$. Additionally, the Pauli gate sampling of NT will act unpredictably on the coherent correction to the Pauli noise $\mathcal{N}^*_c$ from Eq.~(\ref{eq:pauli_noise_finite_sampling}), yielding another correction $\tilde{\mathcal{N}^*_\text{c}}$, whose contribution is \textit{a priori} of small magnitude as well. Thus, we obtain as a final effective noise channel:
\begin{align}
\mathcal{N}_p \xrightarrow[N_\text{NT}\text{ circuits}]{\,\,NT\,\,} \mathcal{N}_d &\equiv (1-\delta_\text{RC}-\delta_\text{NT})\mathcal{N}_d^\infty\notag\\
&+ \delta_\text{NT}\mathcal{N}^*_\text{p}+ \delta_\text{RC}\tilde{\mathcal{N}^*_\text{c}}.
\label{eq:depo_noise_finite_sampling}
\end{align}

Let us now consider the expectation value of an observable measured at the output of the circuit of interest using RC+NT with finite sampling. The average of the outputs of the $N_{\rm NT} = N_{\rm RC}$ circuits must then be multiplied by the sampling factor of the overall circuit, $\gamma^{N_{\rm CNOT}}$ to obtain the expectation value corresponding to the effective target depolarizing noise on all junctions. In the notations of Eq.~\eqref{eq:expectation_value_RC_finite_sampling}, this can be written as

\begin{align}
\langle \hat{O}\rangle_d^\infty  \xrightarrow[N_\text{NT}\text{ circuits}]{\,\,NT\,\,} \langle \hat{O}\rangle_d,
\end{align}
with
\begin{align}
\langle \hat{O}\rangle_d \equiv \langle \hat{O} \rangle_d^\infty  + \gamma^{N_\text{CNOT}}\left[\Delta\langle \hat{O}\rangle_{NT} + \Delta \langle \hat{O} \rangle_c
+ \Delta \langle \hat{O} \rangle_\text{unk.} \right].
\label{eq:depo_observable_finite_sampling}
\end{align}
Here $\Delta\langle \hat{O}\rangle_{\mathrm{NT}}$ is the deviation from the infinite-sampling expectation value $\langle \hat{O}\rangle_d^\infty $ attributed to the finite NT sampling  (it scales as $\delta_\text{NT} \sim 1/\sqrt{N_{\mathrm{NT}}}$).
$\Delta \langle \hat{O} \rangle_c$ is the deviation stemming from the finite RC sampling (i.e., the residual coherent noise) and scales as $\delta_\text{RC} \sim 1/\sqrt{N_\text{RC}}=1/\sqrt{N_{\text{NT}}}$. Just as before, $\Delta \langle \hat{O} \rangle_\text{unk.}$ encompasses the effect of deviations from assumptions (i) and (ii) made in Sec.~\ref{sec:raw_noise} and is a priori unknown. Comparing with Eq.~\eqref{eq:expectation_value_RC_finite_sampling}, one sees that the NT protocol enhances the finite-sampling corrections by a factor $\gamma^{N_{\rm CNOT}}$.

\subsection{Utility of NT: combining with error mitigation}
\label{subsec:NT+error_mitig}

Combining the RC+NT protocol with error mitigation by NEC requires dividing the measured expectation value by NEC-estimated fidelity $\mathcal{F}_\text{NEC}$, cf.~Sec.~\ref{sec:NEC}. Using the notation introduced above, the mitigated result of the RC+NT+NEC protocol performed on a NISQ device with finite sampling can be expressed as

\begin{align}
\langle \hat{O}\rangle_m &\equiv \langle \hat{O} \rangle_m^\infty  + \sigma\left[\Delta\langle \hat{O}\rangle_{NT} + \Delta \langle \hat{O} \rangle_c
+ \Delta \langle \hat{O} \rangle_\text{unk.} \right]\notag\\
&= \langle \hat{O} \rangle_m^\infty  + \Delta\langle \hat{O}\rangle_{NT}' + \Delta \langle \hat{O} \rangle_c'
+ \Delta \langle \hat{O} \rangle_\text{unk.}',
\label{eq:mitigated_observable_finite_sampling}
\end{align}
where $\langle \hat{O} \rangle_m^\infty = \mathcal{F}_\text{NEC}^{-1}\langle \hat{O} \rangle_d^\infty$ is the final error-mitigated expectation value of the observable in the infinite-sampling limit and given assumptions (i) and (ii) from Sec.~\ref{sec:raw_noise} hold valid; the factor $\sigma=\gamma^{N_\text{CNOT}}\mathcal{F}_\text{NEC}^{-1}$, introduced in Sec.~\ref{sec:NEC}, amplifies the inaccuracies, as was discussed in Sec.~\ref{sec:Choosing_optimal_target_noise}. In the second line of the equation, we introduce a notation for the amplified corrections due to finite sampling ($\Delta\langle \hat{O}\rangle_{NT}'$, $\Delta\langle \hat{O}\rangle_c'$) and due to assumption breaking ($\Delta \langle \hat{O} \rangle_\text{unk.}'$).

The potential utility of employing NT is thus a matter of a trade-off. On one hand, employing NT may increase the efficiency of EM. On the other hand it amplifies the errors due to finite sampling and unaccounted noise. In the next two sections, we investigate the extent to which NT can be useful in practice.

\section{Classical emulation of NISQ devices: Accuracy improvement using RC+NT+NEC}
\label{sec:classical_sims}

To assess the efficiency of the NT-improved error mitigation method, we apply it to quantum simulations of a quench in the BCS model. We have previously studied error mitigation in this problem, using RC+NEC \cite{perrin2024}, for $3$ qubits ($3$ Cooper pairs) and deep quantum circuits (long times, over $100$ CNOT gates).

In this section, we classically emulate the performances of a NISQ device, with the noise closely mimicking the noise of an actual NISQ device. We find that the use of NT enables noticeable accuracy improvement for intermediate and long evolution times. Below, we discuss these classical emulations and their results in detail. The results of running the protocol on actual NISQ devices are presented in Sec.~\ref{sec:NISQ}.

\subsection{BCS quench simulation}
\label{subsec:BCS_exp_classical_sims}

In this section, we give a brief summary of the BCS quench simulation protocol used in Ref.~\cite{perrin2024} and the results obtained there. Below, we will use the same BCS quench simulation to benchmark the accuracy of the RC+NT+NEC protocol described in Sec.~\ref{sec:NT_to_depo}, cf.~Fig.~\ref{fig:workflow}.

We consider the BCS Hamiltonian in the spin language,
\begin{align}
H_\text{BCS}&=-\sum_{j=0}^{N_q-1}\left(\epsilon_j -\frac{g}{2}\right)Z_j
-\frac{g}{2}\sum_{ \substack{i,j = 0\\i<j}}^{N_q-1}\left(X_i X_j + Y_i Y_j\right),
\label{eq:BCS_pauli}
\end{align}
with $N_q$ being the number of qubits/Cooper pairs, $X_i$, $Y_i$, $Z_i$ being the Pauli matrices at site $i$, and $\epsilon_j$, $g$ being the model parameters. For the quench simulation, we initialize the qubits in a non-equilibrium product state and simulate the system evolution under $H_\text{BCS}$. The evolution is broken down into Trotter steps that can be implemented on a quantum computer. We measure a set of local (Pauli) observables,
\begin{equation}
\mathcal{D} = \{X_0, Y_1, Z_2, X_0 Y_1, Y_1 Z_2, X_0 Z_2, X_0 Y_1 Z_2 \},
\label{eq:set_of_observables}
\end{equation}
after the evolution of various durations. The measured time-dependent expectation values of the observables are then compared to their theoretical values.

In Ref.~\cite{perrin2024}, we used $N_q = 3$ and up to 15 Trotter steps (resulting in up to 135 CNOT gates in the evolution circuit) on NISQ devices. We achieved a good accuracy of the measured expectation values for short- and intermediate-duration evolution. However, for long evolution circuits, involving over $100$ CNOT gates, the results showed large deviations from the ideal predictions, indicating the limits of the RC+NEC method.

We have attributed this reduction of accuracy for deeper circuits to three main factors: the accumulation of single-qubit-gate errors, the possible temporal correlations of the noise, and the fact that the noise channel of the active qubits of the CNOT gates remains a Pauli noise and not a depolarizing noise, which limits the efficiency of NEC. The first two factors correspond to violations of our assumptions (i) and (ii) in Sec.~\ref{sec:raw_noise}.

The third factor---that the noise has Pauli structure, and does not reduce to the depolarizing noise---can be eliminated using the NT protocol introduced above. Indeed, the NT protocol introduced in Sec.~\ref{sec:NT_to_depo} can convert a local Pauli noise to a local depolarizing noise through appropriate stochastic sampling. This does not fully align the noise structure with that preferred by NEC (global depolarizing noise). Yet, this may lead to an improvement of results. Below we \textit{emulate} the execution of RC+NT+NEC protocol on NISQ devices in order to evaluate the potential for such improvement.

\subsection{Classical emulation of noise}
\label{subsec:noise_in_classical_emulation}

Here we describe the noise channel we implement in our classical emulations of NISQ devices. Up to the subtleties described here below, we model the Pauli noise that corresponds to the actual quantum computers at the time of running our experiments described in Sec.~\ref{sec:NISQ}.

We model the noise of a quantum computer as a Pauli noise channel. Thus, the following trials do not include any coherent part of the noise, $\mathcal{N}_c=0$. This noise is the one that would be obtained on the QC in the infinite-sampling limit of the RC procedure ($\delta_\text{RC}\to 0$ in Eq.~\eqref{eq:pauli_noise_finite_sampling}). Further, our modelling includes no single-qubit or non-Markovian noise (cf.~Sec.~\ref{sec:raw_noise}). Therefore, $\Delta \langle \hat{O} \rangle_c = 0$ and $\Delta\langle \hat{O}\rangle_\text{unk.} = 0$ in our emulations. This enables a direct assessment of the potential accuracy improvement due to using NT, cf.~Eqs.~(\ref{eq:expectation_value_RC_finite_sampling}, \ref{eq:mitigated_observable_finite_sampling}).

In order to make this assessment realistic, we use the Pauli noise extracted from actual NISQ devices by means of PNT. Indeed, our protocol requires the knowledge of the noise parameters in order to use the appropriate sampling distribution in NT, cf.~Fig.~\ref{fig:workflow}. When running our NISQ experiments that are reported in Sec.~\ref{sec:NISQ}, we have extracted the Pauli error rates. It is these rates that we use in our classical emulations. More details about the implementation of noise in our classical emulations are given in Appendix~\ref{app:classical_simulations}.

\subsection{Protocol}
\label{subsec:protocol_classical_sims}

In order to evaluate various aspects of the NT technique, we perform four different trials, which correspond to the classical emulation of the BCS quench simulation with four different noise channels.

\begin{enumerate}
    \item[T1] \textbf{Pauli noise channel, $\mathcal{N}_p^\infty$, cf.~Eq.~\eqref{eq:pauli_noise_RC_infinite_sampling}.} In this trial we use the Pauli noise channel extracted from RC+PNT on \texttt{ibm\_hanoi} and use NEC for error mitigation of the results. This essentially performs the classical emulation of the protocol of Ref.~\cite{perrin2024} and serves as a baseline.

    \item[T2] \textbf{Target depolarizing noise channel of NT, $\mathcal{N}_d^\infty$, cf.~Eq.~\eqref{eq:depo_noise_infinite_sampling}.} We use the depolarizing noise channel with the optimal depolarizing parameter $\lambda_d = \lambda_d^\text{opt}$, chosen (for each 2-qubit junction) according to the procedure outlined after Eq.~\eqref{eq:prefactor_NT_NEC}. Further, using this noise channel assumes the infinite sampling limit in the NT procedure, so that $\delta_\text{NT}\to 0$ in Eq.~\eqref{eq:depo_noise_finite_sampling}. The errors are mitigated using NEC. This trial, therefore, shows the maximal possible accuracy improvement that can be obtained from the use of NT in our protocol.

    \item[T3] \textbf{Target depolarizing noise channel given finite NT sampling, $\mathcal{N}_d$, cf.~Eq.~\eqref{eq:depo_noise_finite_sampling}.} We perform the entire procedure RC+NT+NEC, as described in Fig.~\ref{fig:workflow}, to turn the emulated $\mathcal{N}_p^\infty$ into $\mathcal{N}_d$ with a finite sampling of $N_{\mathrm{NT}} = 10^4$ circuits, realistic for a NISQ experiment. The target depolarizing noise is the same as in T2, and the output is mitigated using NEC as well. T3 aims to mimic what can be expected from a true NISQ experiment, with finite circuit-sampling capabilities.

    \item[T4] \textbf{Average depolarizing noise channel, $\mathcal{N}_{d, \text{avg}}^\infty$.} We assume infinite-sampling limit of NT, yet choose the target depolarizing parameter to be \textit{not} $\lambda_d^\text{opt}$, but simply the average of the Pauli error rates, $\lambda_d = \sum_{i \neq \mathbb{I} \otimes \mathbb{I}} \lambda_i^{(p)}/15$. This trial is designed to evaluate the improvement from changing the noise structure --- without reducing the overall noise strength, which typically happens due to the optimization procedure described below Eq.~\eqref{eq:prefactor_NT_NEC}.
\end{enumerate}

The result of these four trials will allow us to precisely characterize the capability of the NT protocol to boost error mitigation, depending on the available circuit sampling, as well as to understand the effect of changing the noise structure with and without reducing the noise strength.

\subsection{Results}
\label{subsec:results_classical_sims}

\begin{figure*}[t]
    \centering
  \includegraphics[width=0.32\textwidth]{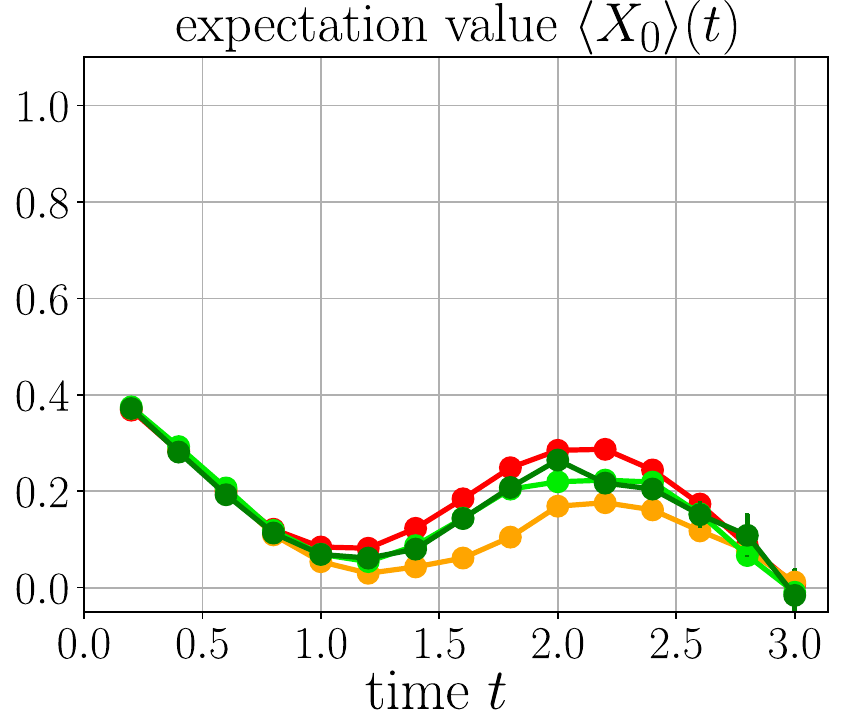}\hfill  \includegraphics[width=0.32\textwidth]{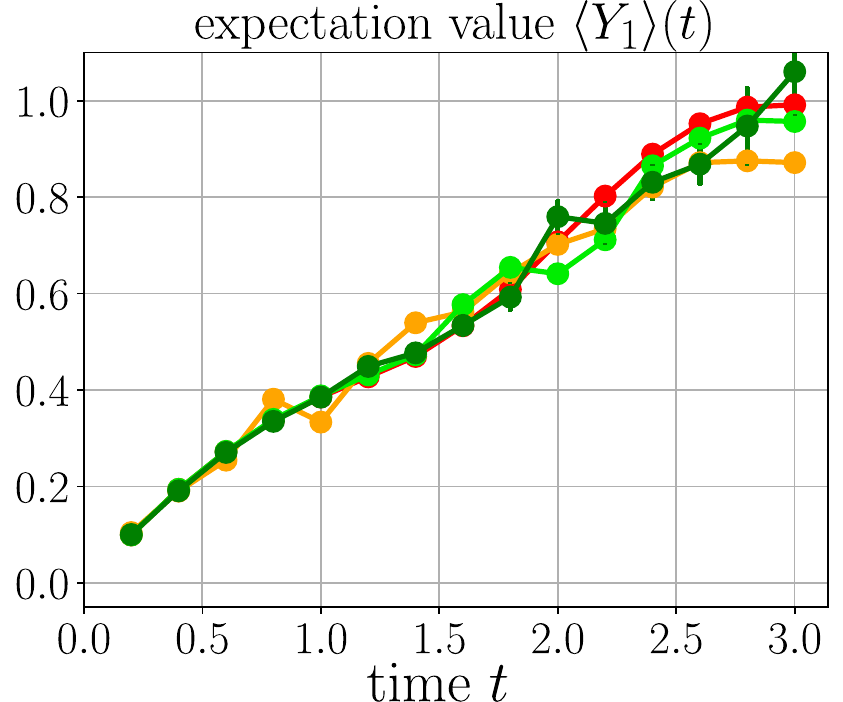}\hfill
  \includegraphics[width=0.32\textwidth]{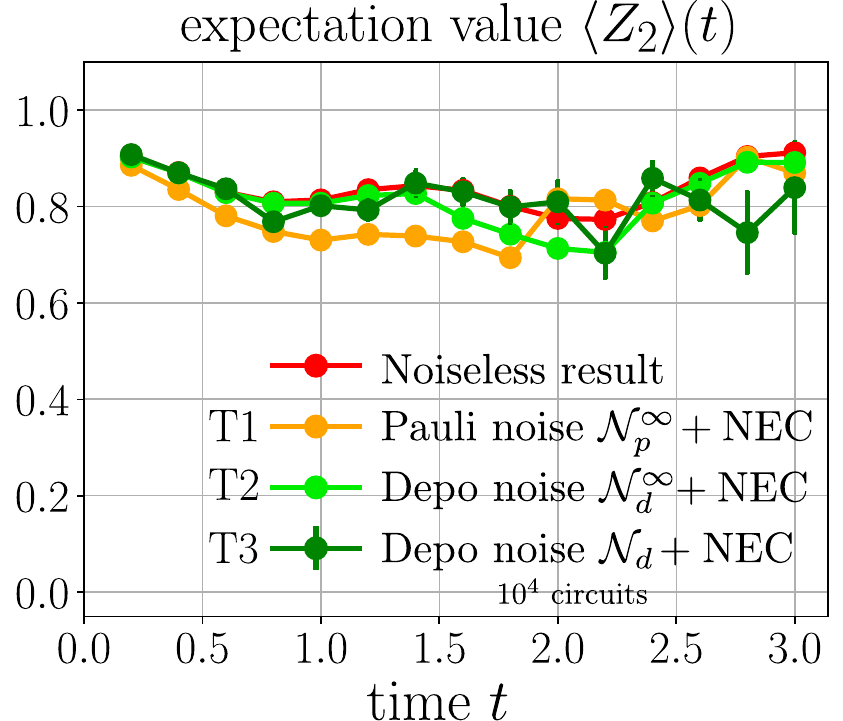}
    \caption{Classical emulation of various error mitigation protocols, as defined in Sec.~\ref{subsec:protocol_classical_sims}. The emulations include the noise that corresponds to the Pauli noise on \texttt{ibm\_hanoi}, cf.~Sec.~\ref{subsec:noise_in_classical_emulation}. We show the evolution of $X_0$, $Y_1$, and $Z_2$ observables over $15$ Trotter steps (the circuits for the longest simulation time contain 135 CNOT gates). The red dots show the perfect result in the noiseless case. The other curves show the results that correspond to trial protocols defined in Sec.~\ref{subsec:protocol_classical_sims}. The orange dots show the result obtained by using NEC directly on a Pauli noise channel $\mathcal{N}_p^\infty$ (T1). The light green dots show the results obtained by using NEC with an optimally selected purely depolarizing channel $\mathcal{N}_d^\infty$ (T2); these results show the potential improvement due to NT in the infinite sampling limit. The dark green dots show the results of the NT protocol using a finite sample size of $N_{\mathrm{NT}} = 10^4$ circuits, which can realistically be run on a NISQ device (T3); error bars indicate the uncertainty associated with both the NT finite sampling and shot noise.}
    \label{fig:time_evol_class_sim}
\end{figure*}

\begin{figure}
    \centering
  \includegraphics[width=\linewidth]{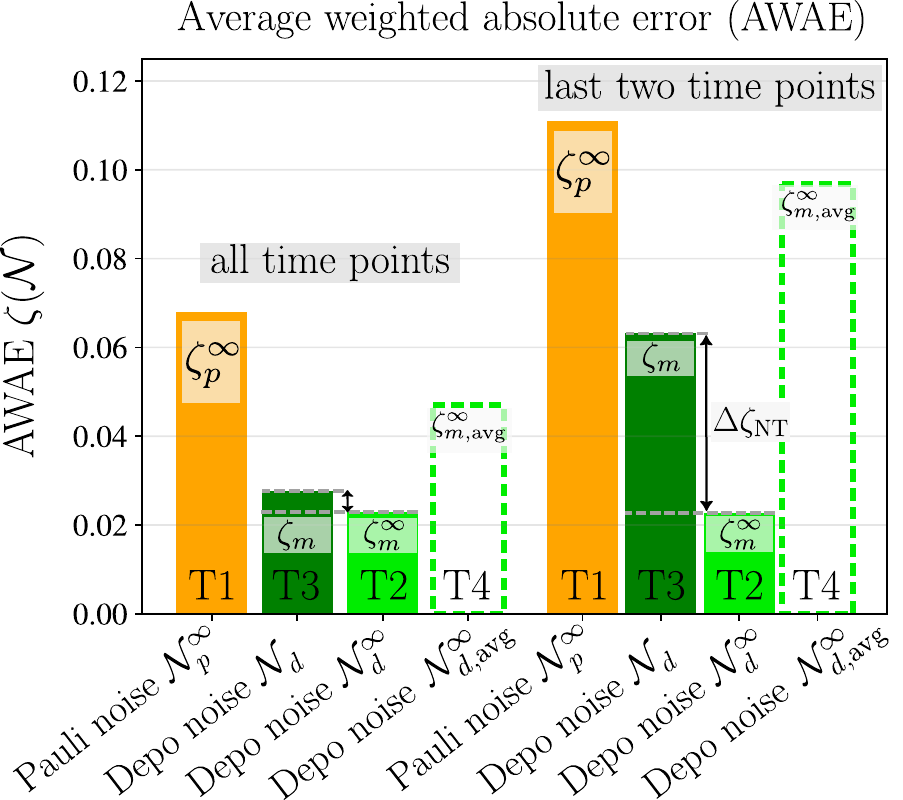}   \\
    \caption{Average weighted absolute error (AWAE) for our classical emulations of the BCS model simulation for the four trial protocols T1--T4 defined in Sec.~\ref{subsec:protocol_classical_sims}. The average is performed over the expectation values of 7 different observables, cf.~Eq.~\eqref{eq:set_of_observables}. The results involve averaging over all time points (left side) and over the last two time points (right side). The bars are labeled with trial numbers T1-T4, and the corresponding noise channels are marked on the $x$-axis. In all cases, the expectation values are error-mitigated using NEC. The contribution to the AWAE due to the finite-sampling of the NT protocol, $\Delta \zeta_{\text{NT}}$, is indicated on the plot by black double arrows. One sees that the use of NT (even in the realistic finite-sampling version, see T3) allows for a substantial reduction of the errors (compare to T1). One further sees that part of the improvement comes purely from changing the noise structure (see T4).}
    \label{fig:error_class_sim}
\end{figure}

The results of the trials defined above are presented in Figs.~\ref{fig:time_evol_class_sim} and \ref{fig:error_class_sim}. Figure~\ref{fig:time_evol_class_sim} shows the time evolution for a few observables from the list in Eq.~\eqref{eq:set_of_observables} under the protocols of T1--T3. The NT tends to improve the results compared to applying pure NEC ($\mathcal{N}_p^\infty$+NEC). This is true both for the idealistic infinite-sampling NT ($\mathcal{N}_d^\infty$+NEC) and the realistic finite-sampling version ($\mathcal{N}_d$+NEC).

This trend is further illustrated in Fig.~\ref{fig:error_class_sim}, which shows the \textit{average weighted absolute error} $\zeta(\mathcal{N})$ (AWAE) for various trials. We define AWAE as follows: For a given trial (which also corresponds to a certain noise channel $\mathcal{N}$), we collect the expectation values of all the observables on the list in Eq.~\eqref{eq:set_of_observables}. We compute the average distance of these expectation values from their noiseless (perfect) values. The AWAE is then given by a weighted average of these distances over all the observables and all time points. The weights are given by the respective noiseless expectation values, which reduces the influence of the observables whose value is close to zero. \footnote{
The AWAE, originally introduced in \cite{perrin2024}, is specifically designed to minimize the influence of the observables whose noiseless expectation value is near $0$. Such observables provide limited insight into the efficiency of our algorithm, as the noise tends to relax the expectation values of observables to zero anyway. Conversely, including such observables in the average would create a false feeling of a better quality of a quantum computer results.
} In other words:
\begin{align}
\zeta(\mathcal{N}) \equiv \frac{1}{Z}\smash{\sum_{ \hat{O}\in\mathcal{D} }}\sum_{j}\abs{\langle \hat{O} \rangle^{\mathrm{perf.}}(t_j)}\abs{\langle \hat{O} \rangle^{(\mathcal{N})}(t_j) -\langle \hat{O} \rangle^{\mathrm{perf.}}(t_j)},
\label{eq:AWAE}
\end{align}
where $\langle \hat{O} \rangle^{\mathrm{perf.}}$ and $\langle \hat{O} \rangle^{(\mathcal{N})}$ denote respectively the noiseless (perfect) result and the noisy (but possibly error-mitigated) result for the same observable; $t_j$ denotes the different time points, and $Z$ is the normalization factor
\begin{align}
Z = \sum_{\hat{O}\in\mathcal{D}}\sum_{j}\abs{\langle \hat{O} \rangle^{\mathrm{perf.}}(t_j)}.
\label{eq:AWAE-NORM}
\end{align}

For brevity, we introduce the notation for AWAE without explicit reference to the noise channel. We attach the identifiers of the noise channel to the AWAE symbol itself, e.g., $\zeta_d^\infty \equiv \zeta(\mathcal{N}_d^\infty)$.

As discussed in the previous section, each mitigated expectation value $\smash{\langle \hat{O} \rangle}$ has corrections due to various sources of errors, which we denoted as $\Delta \langle \hat{O} \rangle'$ (see, Eq.~\eqref{eq:mitigated_observable_finite_sampling}). By extension, we denote the AWAE component due to these corrections as $\Delta\zeta$ (again, keeping the sub/superscritpts of the corresponding noise channel).

In Fig.~\ref{fig:error_class_sim}, we present our results for the AWAE $\zeta$ corresponding to each of the $4$ trials defined in Sec.~\ref{subsec:protocol_classical_sims}. The left part of the figure presents the AWAE where the time average is performed over all time points, $j\in [1,15]$. For the right part of the figure, the time average involves only the last two time points, $j\in [14,15]$, which corresponds to the deepest circuits, for which the effect of noise is the strongest.

We see that the ideal, infinite-sampling [c]RC + [c]NT + NEC method (which produces the effective noise channel $\mathcal{N}_d^\infty$, in light green) indeed yields a significant improvement in accuracy compared to the [c]RC + NEC case (with effective noise channel $\mathcal{N}_p^\infty$, in orange) \footnote{Here by using [c] in the notation, we remind the reader that crosstalk-aware versions of the protocols are used, see the opening of Sec.~\ref{sec:NT_to_depo}. However, we omit the ``c'' in what follows for brevity.}. This trend is also reflected in the finite-sampling version of NT, with effective noise channel $\mathcal{N}_d$ (dark green).

Thus, our classical emulations predict that the NT method should allow for a significant improvement of the results of NISQ devices. The difference between the dark green and light green bars corresponds to the effect of finite sampling in the NT protocol. Using the notation introduced in Sec.~\ref{sec:protocol_on_NISQ}, this quantity corresponds to the correction $\Delta\zeta_\text{NT}\sim\gamma^{N_\text{CNOT}}\mathcal{F}_\text{NEC}^{-1}\Delta\langle \hat{O} \rangle_{NT}$ from Eq.~\eqref{eq:mitigated_observable_finite_sampling}, as indicated on the plot. The other errors from Eq.~\eqref{eq:mitigated_observable_finite_sampling}, $\Delta\langle \hat{O} \rangle_{c}$ and $\Delta\langle \hat{O} \rangle_\text{unk.}$ are absent in our classical emulations, and so are $\Delta\zeta_c$ and $\Delta\zeta_\text{unk.}$.

The rightmost (dashed light green) bars show the result of trial T4. For this trial, the target noise channel $\mathcal{N}_{d,\text{avg}}$ is selected so that the depolarization strength $\lambda_d$ is the average of the Pauli errors $\lambda^{p}_{i}$ in the original noise channel $\mathcal{N}_p^\infty$ (as opposed to selecting $\lambda_d$ in order to minimize the factor $\sigma$ in Eq.~\eqref{eq:prefactor_NT_NEC}, which is done for $\mathcal{N}_d^\infty$). The comparison between T1, T2, and T4 shows that only part of the NT improvement stems from the overall reduction of noise strength; a significant part of the improvement comes from changing the noise structure.

In conclusion, as expected, our classical emulations of a NISQ device show that the NT protocol should yield a significant improvement in accuracy of quantum simulations on such devices. This is conditioned on two assumptions, cf.~Sec.~\ref{sec:raw_noise}: (i) the Markovianity of noise on two-qubit gates and (ii) negligibly small noise of single-qubit gates. The improvement has two origins of comparable importance: overall reduction of the noise strength and changing the noise structure in the NT method. This improvement is substantial not only under pefect execution of the NT protocol, but also when using only a finite number of sampling circuits. On actual NISQ devices, residual coherent noise and breaking of assumptions (i) and (ii) may affect the protocol performance. We present the results of actual NISQ runs in the next section.

\section{Experiments on NISQ}
\label{sec:NISQ}

Having seen the potential for improvement, as evidenced by the classical emulation above, it is important to validate the protocol on an actual NISQ device. In this section, we show our results for the RC + NT + NEC protocol performed on an IBMQ QC, \texttt{ibm\_hanoi}. The same experiments have been performed on another QC, \texttt{ibmq\_ehningen}, and are qualitatively similar, up to an overall lower accuracy; for the sake of clarity, they are presented in Appendix.~\ref{app:ehningen_results}.

In both cases, we find that the NT protocol allows for a practical improvement on the real quantum hardware in principle, with infinite number of NT sampling circuits. However, with the finite sampling we used RC+NT+NEC performs worse than the basic RC+NEC protocol.

At the same time, the comparison of the two protocols allows for gaining advanced insights into the structure of the noise on real quantum computers. We present the details of this analysis in the present section and discuss its potential usefulness for diagnozing quantum computers in Sec.~\ref{sec:NT_for_diagnostics}.

\subsection{Protocol}
\label{sec:exp_NISQ_protocol}

The protocol is the same as the one described in Section~\ref{sec:classical_sims}, trial 3 (T3): the NT protocol is executed with the finite number of sampling circuits, $N_{\text{NT}} = 10^4$, one shot per circuit. We use the crosstalk version of RC+NT+NEC to simulate the \textit{last two} time steps (i.e. quantum circuits with $j = 14$ and $j = 15$ Trotter steps respectively) BCS dynamics on $3$ qubits of \texttt{ibm\_hanoi}.

This machine has a layout with relatively low connectivity, and we choose three qubits on a line to perform our experiment, targeting the junctions in which the CNOT gate error is the lowest in magnitude, identified by performing PNT on all junctions. Having selected the qubits, we determine the target strength of the depolarizing noise, $\lambda_d$, separately for each junction, in order to minimize the error amplification factor $\sigma$ (see Eq.~\eqref{eq:prefactor_NT_NEC}, cf.~Sec.~\ref{subsec:NT+error_mitig}), with the aim of achieving maximally accurate results. We run $N_{\text{NT}} = 10^4$ circuits, using multiple \texttt{qiskit-runtime} sessions \cite{Javadi-Abhari2024} (5 or 6 per single time point $t_j$), which allows to perform the Pauli noise tomography, generate and run the NT circuits in a timely manner. This also enables compensating for the noise evolution on the quantum computers through adjusting the NT in each runtime session. The details are described in Appendices~\ref{app:selection},~\ref{app:runtime}.

In addition, we alleviate the impact of noisy readout by employing a REC scheme based on the iterative Bayesian unfolding method ~\cite{nachman2020}. This protocol requires a prior tomography of the measurement noise channel, but has a negligible sampling overhead compared to the rest of the protocol.

\subsection{Accuracy (non-)improvement using NT+NEC}
\label{sec:accuracy_non-improvement_NT+NEC}

\begin{figure}[h!!]
    \centering
    \includegraphics[width=\linewidth]{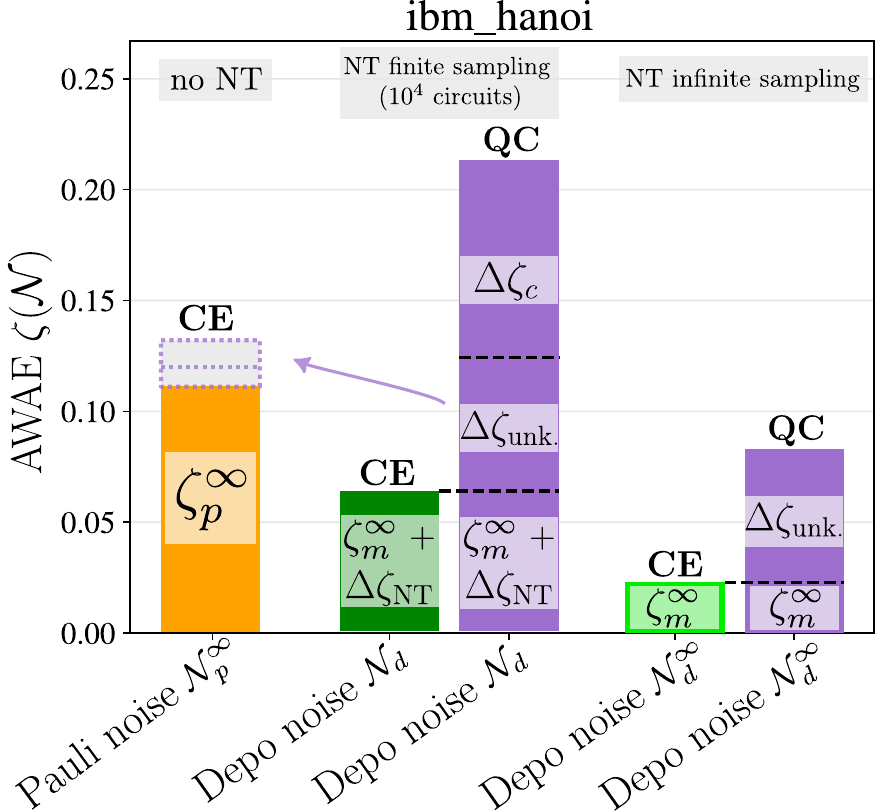}
    \caption{Comparison of the average weighted absolute error $\zeta$ for the last two time points in classical emulations (CE) and in actual quantum computer runs (QC). Purple bars show the QC results obtained on \texttt{ibm\_hanoi}, and the green and the orange bars show the classical emulation results from Fig.~\ref{fig:error_class_sim}. The purple bar in the center shows $\zeta$ obtained in the quantum computer runs with RC+NT+NEC protocol using $N_{\text{NT}} = 10^4$ sampling circuits. Comparing it to the dark-green bar in the center, one identifies the contributions of the residual coherent and unknown noise channels in the QC runs. The dashed purple blocks on top of the orange bar show the estimated contribution of these noise sources in the RC+NEC protocol, see Sec.~\ref{sec:accuracy_non-improvement_NT+NEC} for details. The purple bar on the right shows an estimate for AWAE in the limit of $N_{\text{NT}} = \infty$, which eliminates the residual coherent noise completely; the extrapolation procedure producing this estimate is described in Sec.~\ref{sec:extrapolation}. Overall, the application of NT does not produce an accuracy improvement when using $10^4$ sampling circuits due to the amplification of non-Pauli noise sources by the NT protocol. However, eliminating the residual coherent noise by increasing the number of sampling circuits can provide improvement compared to the bare RC+NEC protocol. Lastly, comparing the two protocols provides a diagnostic for various sources of noise present on a quantum computer, as discussed in Sec.~\ref{sec:NT_for_diagnostics}.}
    \label{fig:error hanoi}
\end{figure}

The QC results for \texttt{ibm\_hanoi} are summarized by the center purple bar in Fig.~\ref{fig:error hanoi}, which shows the AWAE $\zeta(\mathcal{N})$, cf.~Eqs.~(\ref{eq:AWAE}, \ref{eq:AWAE-NORM}). The averaging of $\zeta$ has been performed over $7$ observables (see Eq.~\eqref{eq:set_of_observables}) and the last two time points ($j=14, 15$ or $t_j = 2.8, 3.0$). To the left, we show the predictions of the classical emulations for the RC+NT+NEC protocol (dark-green bar) and for the RC+NEC protocol (orange bar) \cite{foot_no_quantum_results}. These are the same values as in the right part of Fig.~\ref{fig:error_class_sim}.

Remarkably, the error of the computation on the actual quantum computer is significantly bigger than that predicted by classical emulation (purple bar vs dark-green bar in the middle of Fig.~\ref{fig:error hanoi}). We remind the reader that the classical emulations include \textit{the very same Pauli noise} as extracted by PNT during the runs on the quantum computer. Moreover, the classical emulation takes into account all the practical nitty-gritty, such as multiple runtime sessions and different noise in each of those. This forces one to conclude that other sources of noise are present on the quantum computer. These can be the residual coherent noise of CNOT gates (due to finite RC sampling), non-Markovian noise or single-qubit noise. The latter two options correspond to violating assumptions (i) and (ii) from Sec.~\ref{sec:raw_noise}.
One can actually disentangle the contributions of the extra noise sources through an in-depth analysis of our data. The result of this analysis is marked in Fig.~\ref{fig:error hanoi}. $\Delta \zeta_c$ represents the contribution  to $\zeta$ of the residual coherent noise, while $\Delta \zeta_\text{unk.}$ stands for the contribution due to the violation of assumptions (i) and (ii). Before we present this analysis in Sec.~\ref{sec:extrapolation}, let us address a simpler question.

Do the RC+NT+NEC results on a quantum computer outperform the basic RC+NEC protocol we used in Ref.~\cite{perrin2024}? Answering this question would have been easy with the data for the two protocols taken simultaneously. Unfortunately, as explained in endnote \cite{foot_no_quantum_results}, we do not possess such data and have to make conclusions based on classical emulations and understanding the two protocols.

Nevertheless, the question can be partially answered by a careful analysis of the available data, as we explain now. In Sec.~\ref{sec:protocol_on_NISQ}, we have seen that the contribution of finite-size-sampling corrections to the error of a given observable on the quantum computer run is given by $\Delta\langle \hat{O}\rangle_{NT}' + \Delta \langle \hat{O} \rangle_c' + \Delta \langle \hat{O} \rangle_\text{unk.}'$, cf.~Eq.~\eqref{eq:mitigated_observable_finite_sampling}. This translates to the AWAE components $\Delta\zeta_{NT} + \Delta\zeta_c + \Delta\zeta_\text{unk.}$ forming the purple bar in the center of Fig.~\ref{fig:error hanoi}. On the other hand, our classical emulation of the RC+NT+NEC protocol does not contain any coherent noise, nor the noise violating assumptions (i) and (ii). Therefore, the average weighted absolute error $\zeta$ coming from the classical emulation is purely due to the finite sampling in the NT protocol, $\Delta \zeta_{NT}$. Thus, the absolute difference between the purple and dark-green bars in the middle of Fig.~\ref{fig:error hanoi} yields an estimate of $\Delta\zeta_c + \Delta\zeta_\text{unk.}$.

In the RT+NEC protocol, the correction $\Delta\zeta_{NT}$ would naturally be absent. However, what would be the effect on $\zeta$ of the extra noise sources leading to $\Delta \langle \hat{O} \rangle_c' + \Delta \langle \hat{O} \rangle_\text{unk.}'$? In the RC+NT+NEC protocol, the errors are amplified by the factor $\sigma = \gamma^{N_\text{CNOT}}/\mathcal{F}_\text{NEC}$, cf.~Eq.~\eqref{eq:mitigated_observable_finite_sampling}. In the absence of NT, the error amplification factor is just $1/\mathcal{F}_\text{NEC}$. The $\mathcal{F}_\text{NEC}$ for two protocols is, however, not the same, as the noise channel is different. Therefore, one can estimate the effect of the extra noise sources as
\begin{align}
\frac{\Delta \langle \hat{O} \rangle_c + \Delta \langle \hat{O} \rangle_\text{unk.}}{\mathcal{F}_\text{NEC}(\mathcal{N}_p^\infty)} = \frac{\Delta \langle \hat{O} \rangle_c' + \Delta \langle \hat{O} \rangle_\text{unk.}'}{\mathcal{F}_\text{NEC}(\mathcal{N}_p^\infty)} \frac{\mathcal{F}_\text{NEC}(\mathcal{N}_d^\infty)}{\gamma^{N_\text{CNOT}}}.
\end{align}

Translated in terms of the AWAE $\zeta$, this yields the estimate shown on top of the orange bar in Fig.~\ref{fig:error hanoi}. Of course, the estimate includes the assumption that the non-application of NT does not significantly change the magnitude of $\Delta \langle \hat{O} \rangle_c + \Delta \langle \hat{O} \rangle_\text{unk.}$, which is reasonable because these are device-bound properties. Our estimate thus shows that under realistic conditions the basic RC+NEC protocol outperforms the RC+NT+NEC protocol. And this is due to the amplification of coherent noise, non-Markovian noise, and the noise associated with single-qubit gates by the finite-sampling NT protocol.

One outstanding question is whether the infinite-sampling NT protocol would enable outperforming the RC+NEC protocol. We answer this question positively and discriminate the contributions of $\Delta \zeta_c$ and $\Delta \zeta_\text{unk.}$ in Sec.~\ref{sec:extrapolation}.

\subsection{Extrapolation to infinite sampling}
\label{sec:extrapolation}

We have seen above that the inclusion of NT in the RC+NT+NEC protocol, though useful according to classical emulations, does not bring the desired accuracy improvement on an actual quantum computer. The reason is that the NT protocol amplifies the unaccounted errors by a big factor $\sigma$, cf.~Eqs.~(\ref{eq:prefactor_NT_NEC}, \ref{eq:mitigated_observable_finite_sampling}). Thus, the practical accuracy improvement is a matter of trade-off between the result improvement through aligning the known noise with the mitigation technique via the NT and the result worsening due to the amplified contribution of the unaccounted noise.

Two of the error types in Eq.~(\ref{eq:mitigated_observable_finite_sampling}) can be completely eliminated through increasing the number of sampling circuits in the RC+NT+NEC protocols. The residual coherent noise $\Delta \langle \hat{O} \rangle_c'$ is present because of imperfect sampling in the RC part of the protocol. And the error in aligning the Pauli noise with the depolarizing noise $\Delta\langle \hat{O}\rangle_{NT}'$ is only present if the NT sampling is imperfect.

It is, therefore, of interest to estimate the would-be AWAE $\zeta$ for the infinite sampling limit in the RC+NT+NEC protocol. This would clarify whether the RC+NT+NEC protocol could be beneficial in principle compared to RC+NEC on \texttt{ibm\_hanoi}.

\begin{figure}[h]
    \centering
    \includegraphics[scale=0.7]{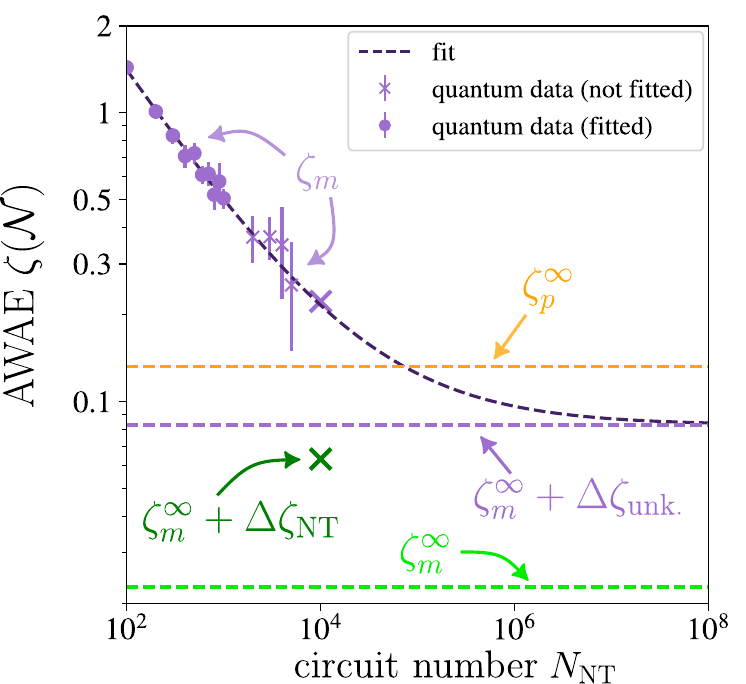}
    \caption{Extrapolation the AWAE $\zeta$ of quantum computer runs on \texttt{ibm\_hanoi} to the infinite-sampling limit. The purple dots and crosses show the $\zeta$ levels for various number of sampling circuits $N_{\text{NT}}$. These $\zeta$ values and error bars are estimated through bootstrapping. The dashed black curve shows the result of fitting the data points with Eq.~\eqref{eq:AWAE_fit_formula}. The horizontal purple dashed line shows the extrapolated $\zeta$ level for infinite sampling, i.e., $b$ from Eq.~\eqref{eq:AWAE_fit_formula}. Other $\zeta$ levels from Fig.~\ref{fig:error hanoi} are shown for reference. The extrapolation shows that given sufficient sampling, the NT protocol can improve the results over the basic RC+NEC protocol not only in classical emulations (horizontal light-green dashed line), but also on a real quantum computer (horizontal purple dashed line).}
    \label{fig:scaling}
\end{figure}

While we cannot perform actual infinite-sampling runs on a quantum computer, it is nevertheless possible to extrapolate the expected AWAE based on the data we have. Below we describe how we do it using bootstrapping. Each of our BCS simulations contains the results for the AWAE $\zeta$ from $10^4$ distinct sampling circuits. We break these down into batches of $10^2$ circuits. We then compute $\zeta$ for each of those batches and  estimate the average $\zeta$ for $100$-circuit runs. By merging the pairs of consecutive batches (first with second, third with fourth etc.), we estimate $\zeta$ for $200$-circuit runs. Similarly, we estimate $\zeta$ for would-be runs with $300$, $400$, ..., $1000$ sampling circuits (solid purple circles in Fig.~\ref{fig:scaling}) and $2000$, $3000$, $4000$, $5000$ sampling circuits (thin purple crosses in Fig.~\ref{fig:scaling}). The above $\zeta$ estimates come with the statistical error bars, as different batches produce different $\zeta$ values. The thick purple cross corresponds to $\zeta$ in our experiment with $10^4$ sampling circuits, same as the purple bar in the middle of Fig.~\ref{fig:error hanoi}.

We expect $\zeta$ to depend on the number of sampling circuits $N_{\text{NT}}$ as
\begin{equation}
f(N_{\text{NT}}) = a/\sqrt{N_{\text{NT}}}+b.
\label{eq:AWAE_fit_formula}
\end{equation}
The first term represents the contribution of the Monte-Carlo sampling corrections $\Delta \zeta_c'$ and $\Delta\zeta_{\text{NT}}'$ to $\zeta$. The second term stands for the contribution to $\zeta$ of the error that does not depend on $N_{\text{NT}}$, $\Delta \zeta_\text{unk.}'$. We fit the above estimates for $\zeta$ with $f(N_{\text{NT}})$ using $a$ and $b$ as fitting parameters. We only use $\zeta$ estimates for up to $1000$-circuit-strong runs. This is because $\zeta$ estimates for larger numbers of circuits rely on very few batches, and thus are prone to statistical noise. \footnote{The AWAE for $10^4$ circuits is a single data point, whose representativity is not a priori clear. The estimates for $2000$-$5000$ circuits are based on averaging $2$-$5$ individual $\zeta$ values. These estimates can thus fluctuate a lot as evidenced by the error bars in Fig.~\ref{fig:error hanoi}.}

The fitted $f(N_{\text{NT}})$ is shown in Fig.~\ref{fig:scaling}. The fitted value of $b$ is shown as the horizontal purple dashed line; this is our estimate for $\zeta$ with an infinite number of sampling circuits, presented as the right purple bar in Fig.~\ref{fig:error hanoi}. Achieving that accuracy, however, requires $N_{\textrm{NT}}\gtrsim10^6$.
We thus conclude that, in principle, the RC+NT+NEC protocol is advantageous to RC+NEC protocol, if a sufficiently high number of sampling circuits $N_{NT}$ can be used. The advantage, though, hardly justifies the additional computation time required by the NT method. At the same time, the above analysis is valuable for performing diagnostics of quantum computers. We explain this in detail in the next section.

\section{DISCUSSION}
\label{sec:Discussion}

\subsection{The value of NT for diagnozing quantum computers}
\label{sec:NT_for_diagnostics}

The above results show limited applicability of the NT method in practice: amplification of unaccounted error sources, inevitably present on a quantum computer, makes the potential improvement insignificant and the required sampling overhead immense. The same amplification, however, makes NT a useful diagnostic tool for understanding the noise of quantum computers. We detail this idea in this section.

Consider the results in Fig.~\ref{fig:error hanoi}. The classical emulation result for infinite sampling (light-green bar) shows the best possible result one can expect from the RC+NT+NEC protocol. The error $\zeta_m^\infty$ stems from the fact that even perfect NT does not align the noise with a \textit{global} depolarizing noise (which is an assumption of the NEC technique). This is the best one can expect from the RC+NT+NEC protocol given the input Pauli noise we extracted on \texttt{ibm\_hanoi}.

The AWAE $\zeta$ of QC runs extrapolated to the infinite-sampling limit (purple bar on the right) is directly related to the corrections $\Delta \langle \hat{O} \rangle_m^\infty + \Delta \langle \hat{O} \rangle_\text{unk.}'$. The second term represents unaccounted error sources (non-Markovian noise and the noise of single-qubit gates).

The $\zeta$ of finite-sampling classical emulations (dark-green bar) accounts for $\Delta \langle \hat{O} \rangle_m^\infty + \Delta\langle \hat{O}\rangle_{NT}'$. The $\zeta$ of QC runs is further increased by the contribution to $\zeta$ of $\Delta \langle \hat{O} \rangle_c' + \Delta \langle \hat{O} \rangle_\text{unk.}'$. Given the knowledge of the contribution of $\Delta \langle \hat{O} \rangle_\text{unk.}'$, one extracts the contribution of $\Delta \langle \hat{O} \rangle_c'$, i.e. the error due to the residual coherent noise, left after finite-sampling RC.

In the absence of NT, the contributions of $\Delta \langle \hat{O} \rangle_c'$ and $\Delta \langle \hat{O} \rangle_\text{unk.}'$ should be scaled down by the factor $\sigma$ (Eq.~(\ref{eq:prefactor_NT_NEC})). They are hardly noticeable in the RC+NEC results, as evidenced by the dashed purple rectangles on top of the orange bar in Fig.~\ref{fig:error hanoi}.

However, improvement of quantum computers requires eliminating even small errors in order to go significantly below the threshold of quantum error correction and achieve fault-tolerant quantum computing (FTQC). In this respect, understanding the types and sources of different errors on a quantum computer is important in order to inform the work on hardware improvement. The above analysis based on the comparison of classical emulation and quantum runs of both RC+NEC and RC+NT+NEC protocols can thus be a useful diagnostic tool.

This capability positions our NT approach as a potentially valuable diagnostic and benchmarking tool for characterizing NISQ devices \cite{Erhard2019, bultrini2023, Cirstoiu2023}, a challenge notoriously hard in the presence of correlated noise \cite{Figueroa-Romero2021, Harper2023}, and provide hardware developers and quantum algorithm designers with detailed insights into a given device's performance at a given time.

\subsection{Compatibility of NT with multiple error mitigation schemes}
\label{sec:NT_compatibility}

In this work, we have only considered the NEC error mitigation scheme, which achieves its maximal efficiency for a depolarizing noise structure~\cite{urbanek2021,perrin2025}. However, the advantage of targeting depolarizing noise ---which can be created with NT--- extends far beyond NEC to virtually all major error mitigation techniques.

Our work stands in interesting contrast to recent advances in error amplification techniques. IBM's landmark paper \cite{kim2023} introduced Probabilistic Error Amplification (PEA), which amplifies noise magnitude while learning the randomly compiled noise model of each junction. However, PEA preserves the noise structure. Our NT approach is fundamentally different: rather than merely scaling noise strength, we actively reshape the noise structure itself.

This distinction is crucial: when noise is rendered effectively gate-independent and depolarizing, simple linear or polynomial zero-noise extrapolation becomes both accurate and sample-efficient \cite{giurgica-tiron2020,he2020}. More generally, methods that actively uniformize noise across qubit junctions consistently enhance the performance of error-mitigation protocols \cite{Majumdar2023,Hour2024}.

Similarly, Virtual Distillation \cite{Koczor2021, Huggins2021} explicitly employs depolarizing channels in its theoretical framework and achieves exponential error suppression specifically under this noise model. The technique's robustness depends critically on the noise maintaining the dominant eigenvector structure, which is naturally preserved in depolarizing channels \cite{Li2023, Vikstal2024, Xu2024a, Bako2025}. Other advanced techniques including Clifford Data Regression and Symmetry Verification also benefit substantially from the structural simplicity of depolarizing noise \cite{bultrini2023, Cai2021}.

Importantly, the NT's ability to engineer quasi-exact depolarizing noise enables quantum error mitigation experiments that are much closer to analytical approaches. The mathematical tractability of depolarizing channels—characterized by a single parameter and symmetric error distribution—allows for precise theoretical predictions and rigorous benchmarking against analytical models \cite{temme2017, Endo2018}. This stands in stark contrast to the complex, multi-parameter noise models typically encountered in real hardware, where analytical treatment becomes intractable. By providing access to this simplified yet realistic noise regime, the NT opens new possibilities for systematic studies of error mitigation performance and enables direct validation of theoretical predictions under controlled conditions.

\subsection{The usefulness of NT for simulating open quantum systems}

Another particularly promising direction for future research is leveraging the inherent noise of NISQ devices as a resource for simulating open quantum system dynamics \cite{Bertrand2025, Swain2025}. When the noise of the quantum computer and the target open system are of comparable strength, our NT method could be employed to align the hardware noise with the desired environmental model without requiring prohibitive sampling overhead. While several studies have begun exploring this concept for specific noise structures \cite{guimaraes2023, Ma2024} or specific classes of systems \cite{Papic2025}, our NT procedure provides a general framework that could be extended beyond Pauli noise models. This approach could transform quantum simulation of dissipative systems, enabling direct hardware-based studies of decoherence, thermalization, and other open-system phenomena using the quantum device's intrinsic noise as a computational resource rather than an obstacle.

\section{Conclusion}
\label{sec:conclusion}

In this work, we have proposed a novel NT technique. The technique takes its roots in the well-known PER/PEC techniques, yet focuses on modifying the noise structure rather than changing its magnitude.

We have investigated the NT's potential for improving the results of quantum computations via aligning the quantum computer's noise with the structure favored by an error mitigation scheme. While our classical emulations predicted significant improvement due to the use of NT, the actual performance on the quantum computers is disappointing. The reason is that the NT amplifies the errors from unaccounted error sources.

This very problem, however, makes the NT a valuable tool for diagnozing quantum computers. Indeed, we have shown that a careful analysis of the protocol data enables insights and quantitative characterization of various small error sources on quantum devices.

Our work thus establishes both a theoretical framework and a diagnostic tool that will become increasingly valuable as quantum computers mature. The fundamental insight—--that noise structure matters as much as noise strength---opens new directions for extracting quantum advantage from noisy devices. As we transition toward the era of early fault-tolerant quantum computing, techniques that can both diagnose and reshape noise will be essential components of the quantum computing toolkit.

\section*{Acknowledgements}

We acknowledge funding from the state of Baden-Württemberg through the Kompetenzzentrum Quantum Computing (Project QC4BW). This work is part of HQI (\url{www.hqi.fr}) and BACQ initiatives and is supported by France 2030 program under the French National Research Agency grants with numbers ANR-22-PNCQ-0002 and ANR-22-QMET-0002 (MetriQs-France). TS is grateful to Departamento de Física, FCFM, University of Chile (Santiago) for hospitality during the ﬁnal stage of this work. HP acknowledges support from the Horizon Europe program HORIZON-CL4-2021-DIGITAL-EMERGING-01-30 via the project 101070144 (EuRyQa).

\bibliography{reference_v2,extras}

\newpage

\appendix

\section{Noise-tailoring protocol}
\label{app:NT}

\par In this Appendix, we outline the theory of NT for the tailoring of a given Pauli noise channel (which can be obtained in practice using randomized compiling \cite{wallman2016,kern2005,hashim2021,cai2019,perrin2024}) to any effective target Pauli noise channel. NT generalizes PEC~\cite{vandenberg2023,temme2017} and PER~\cite{mari2021,mcdonough2022}.

We assume that single-qubit gates are noiseless, and only consider noise affecting two-qubit gates. We start by reminding the reader several approaches to describing noise mathematically, and then discuss the implementation of NT in the absence and presence of crosstalk. In the following, the CNOT gate, native to IBM machines is taken as the $2$-qubit gate of interest, but the discussion remains valid for any type of noisy entangling gate.

\subsection{Pauli transfer matrix and $\chi$ matrix representation}
\label{app:noise_representation}
The impact of noise on quantum gates is conveniently modeled by a complete positive trace-preserving (CPTP) map denoted as $\mathcal{E}$, acting on the space of density matrices. In the context of digital QCs, such a map describing noise is referred to as ``noise channel".

There are various representations of such maps. In this Appendix, we will employ two of them: the Pauli transfer matrix (PTM) representation and the $\chi$-matrix representation, both of which we introduce here.

A general $q$-qubit Pauli noise channel, admits the following representation:
\begin{align}
\mathcal{E}_{q}(\rho)=\sum_{a,b=0}^{4^{q}} \chi_{a,b}^{(q)}
    P_a^{(q)}\,\rho\,P_b^{(q)},\label{eq:pauli_noise_general}
\end{align}
where we introduced the Pauli string operators $P_a^{(q)}\in\{\mathbb{I},X, Y, Z\}^{\otimes q}$ ($X$, $Y$ and $Z$ corresponding to the single-qubit Pauli gates) and the $\chi$-matrix. The probability conservation implies $\sum_a \chi_{a,a}^{(q)} = 1$.

Pauli noise is defined by the requirement that the same Pauli matrix acts from the left and from the right: $\chi_{a,b}^{(q)}=p_a \delta_{ab}$. Then
\begin{align}
\mathcal{E}^{\text{pauli}}_{q}(\rho)=\sum_{a=0}^{4^{q}} p_{a}
    P_a^{(q)}\,\rho\,P_a^{(q)},\label{eq:pauli_noise}
\end{align}
and $\sum_{a}p_{a}=1$. Each term in the sum represents the probability that the corresponding Pauli error occurs on a single run (shot) of the circuit on the QC. The off-diagonal elements $\chi_{a,b}^{(q)}$, $a\neq b$ are called coherences and are associated with coherent noise; they vanish in the case of a Pauli noise channel.

In spite of being rather intuitive, the $\chi$-matrix representation may not be optimal for analytical approaches. To enable the use of linear algebra techniques, one promotes the density matrix $\rho$ to a vector of a new Hilbert space of dimension $4^{q}$, and denotes it as
\begin{align}
 \ket{\rho}\rangle=\left[
        \cdots \rho_a\cdots
    \right]^T\nonumber,
\end{align}
where $\rho_a=\text{Tr}(P_a^{(q)}\,\rho)$. In this representation, noise channels (which are superoperators) become operators that act on the vectorized density matrices from the left side. They are called Pauli Transfer Matrices (PTM) and are defined element-wise as follows:
\begin{equation}
    E^{(q)}_{b,a}=\frac{1}{2^{q}}\text{Tr}(P_b^{(q)}\,\mathcal{E}(P_a^{(q)}))
\end{equation}
The trace-preserving condition sets the first line of this matrix to $(1,0,\cdots,0)$.

For Pauli noise channels, $\smash{E^{(q)}_{b,a}}$ are diagonal matrices whose elements are called the fidelities $\smash{f_a=E_{a,a}^{(q)}}$. The mapping between the $\chi$-matrix and the PTM representation of a Pauli noise channel is done by the  Walsh-Hadamard transformation:
\begin{equation}
   p_a=\frac{1}{4^{q}}\sum_{b=0}^{4^{q}}(-1)^{\langle P_a^{(q)}, P_b^{(q)} \rangle_\text{sp}} f_b,
   \label{eq:WH}
\end{equation}
where $\langle P_a^{(q)}, P_b^{(q)}\rangle_\text{sp}$ denotes the simplectic inner product of Pauli string operators $P_a^{(q)}$ and $P_b^{(q)}$, which vanishes if the Pauli strings commute and equals one otherwise.

\subsection{NT with no crosstalk}

Let us first assume that the quantum device is crosstalk-free, such that the noise channels associated with two-qubit gates act only on the two qubits where the gate is applied. The following NT procedure focuses on the two qubits of a single junction, such that we work with noise channels with $q=2$. The procedure then naturally generalizes to each and all qubit junctions involved in the circuit of interest.

Below we start with a Pauli noise channel present on the device and convert it to the target Pauli channel. We remind the reader that the bare noise channel on a QC is not necessarily of Pauli type. In order to effectively implement a Pauli noise channel on a real QC, one needs to apply the RC method~\cite{wallman2016,kern2005,hashim2021,cai2019}, which consists in randomly dressing CNOT gates with $2$-qubit Pauli strings, while ensuring their overall action remains logically equivalent to the original CNOT. By averaging the circuit outcomes over a sufficient number of randomized versions of the original circuits, the method eliminates coherences of the experimental noise channel, reducing the CNOT gate noise channel to an effective Pauli noise channel. This is the starting point for NT.

\subsubsection{NT towards a target Pauli noise channel}

Assume the knowledge of the original noise channel affecting the CNOT gate on a particular qubit junction. Namely, its PTM representation $E^\text{gate}_2$. In this work, we have employed PNT (more precisely its crosstalk version) systematically and on every qubit junction to obtain a reliable estimate of the coefficients of $E^\text{gate}$ across the quantum device. The PNT procedure and its crosstalk version are described in details in Appendix~\ref{app:PNT}.

\begin{figure}[h!]
    \centering
     \includegraphics[trim={0 2cm 0 1cm},clip,width=\linewidth]{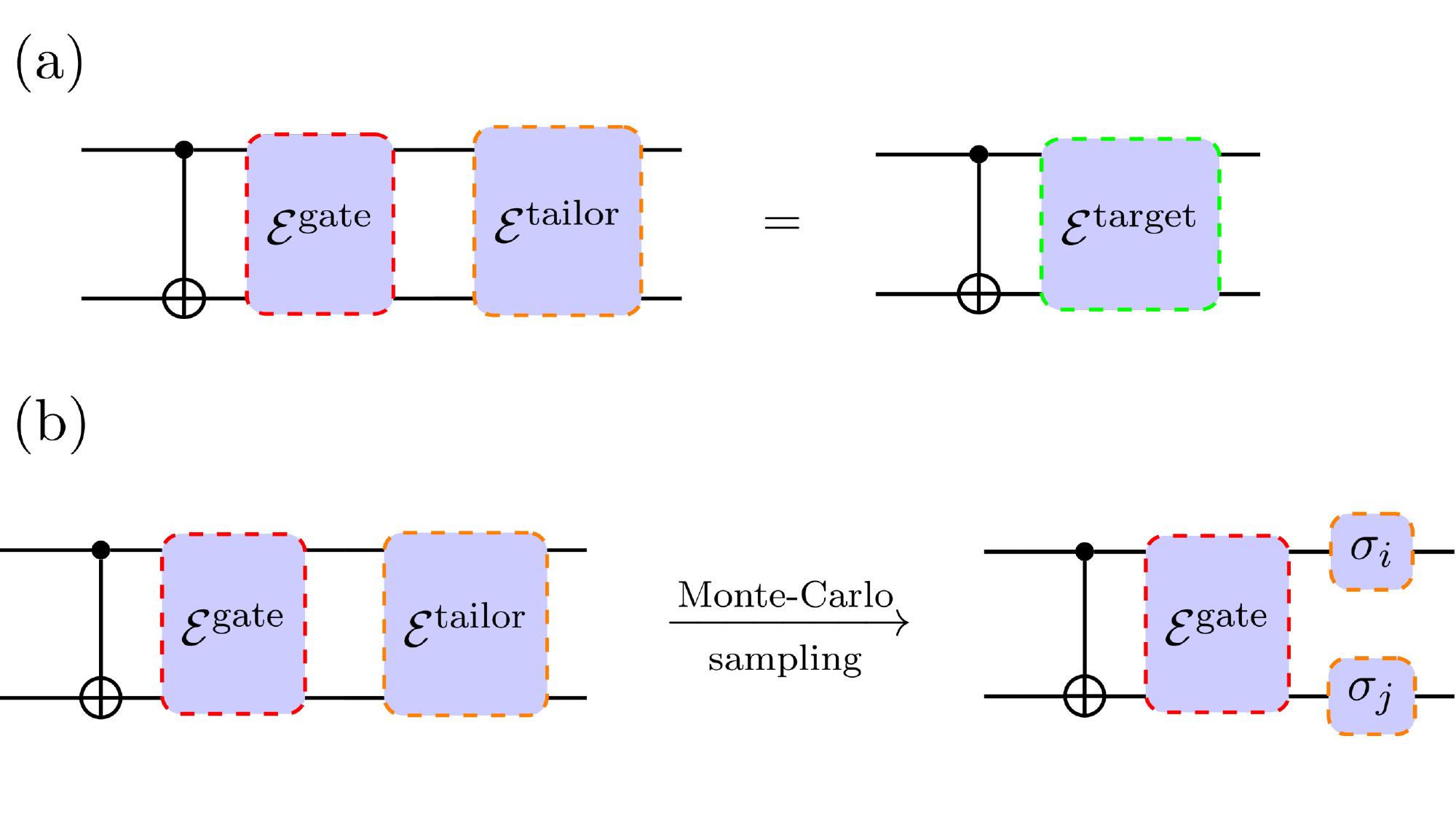}
  \caption{(a) Circuit representation of Eq.~\eqref{eq:noise_target}. (b) Illustration of the implementation of a stochastic noise channel $\mathcal{E}^\text{tailor}_2$ following the application of a noisy CNOT gate with noise channel $\mathcal{E}^\text{gate}_2$. The Monte Carlo sampling is performed over circuits with extra Pauli (or Identity) gates $\sigma_i$ and $\sigma_j$ following the CNOT, randomly sampled following the distribution (\ref{eq:proba}).}
  \label{fig:PEC_noisy_CNOT}
\end{figure}

To effectively tailor $E^\text{gate}_2$ to another target Pauli noise channel $E^\text{target}$, it is sufficient to apply an additional noise channel $E^\text{tailor}_2$ after the application of the two-qubit gate (see Figure~\ref{fig:PEC_noisy_CNOT}). This noise channel should satisfy:
\begin{equation}
E^\text{tailor}_2= E^\text{target}_2(E^\text{gate}_2)^{-1}.
\label{eq:noise_target}
\end{equation}
\par To do so, we randomly implement $2$-qubit Pauli strings $P_a^{(2)}$ on the active qubits of every CNOT gate applied on the junction, according to a certain probability distribution ${p_a}$. The probabilities $p_a$ for each Pauli string are obtained by applying the Walsh-Hadamard transformation to the above equation, mapping the PTM representation to its $\chi$-matrix representation, cf.~Eqs.~(\ref{eq:pauli_noise}--\ref{eq:WH}). Then, $\mathcal{E}^\text{tailor}$ can be effectively implemented by sampling the associated Pauli strings with their respective probabilities.

However, the matrix inversion performed in Eq.~\eqref{eq:noise_target} to obtain $\mathcal{E}^\text{tailor}$ together with the Walsh-Hadamard transform \eqref{eq:WH} usually leads to some $p_a$ being negative. Therefore, we denote the result of the Walsh-Hadamard transform $q_a$, a quasiprobability distribution. In practice, to sample from the quasi-distribution $\{q_a\}$, a mapping to a true probability distribution $\{p_a\}$ can be achieved through the following procedure, for each $a$ one defines:

\begin{align}
    p_{a}=\frac{|q_{a}|}{\gamma},
    \label{eq:proba}
\end{align}
with
\begin{align}
    \gamma=\sum_a |q_{a}|,
    \label{eq:quasiproba}
\end{align}
so that all $p_a \geq 0$ and $\sum_a p_a = 1$. The original quasiprobability distribution can be reconstructed as $q_a=\gamma\ \text{sgn}(q_a) p_a$.

The transformation of signed $q_a$ to $p_a \geq 0$, however, implies an increased cost of the Monte Carlo sampling. Indeed, whenever the stochastic noise channel $\mathcal{E}^\text{tailor}_2$ is implemented by sampling over $p_a$, the quantum circuit outcomes must be multiplied by the sign of the corresponding quasi-probability $q_a$, their sum must be multiplied by the factor $\gamma$. This mutliplies the variance of the associated unbiased estimator by $\gamma$. As a result, achieving a fixed accuracy requires $\gamma$ times more sampling. This is similar to the well-known \textit{sign problem}, common in the Monte-Carlo-based simulations of correlated fermionic systems. Using the normalization $\sum_i q_i=1$, the sampling factor $\gamma$ can be recast as:
\begin{equation}
    \gamma=1+2\sum_{q_i<0}|q_i| > 1,
    \label{eq:gamma}
\end{equation}
which highlights that the sampling cost is directly related to the number and the strength of the negative coefficients.
For a quantum circuit of interest containing $N_{\text{noisy}}$ noisy gates,  (assuming, for simplicity, that each of them is affected by the same Pauli noise channel) the overall sampling cost $\sigma$ for the effective mapping of all channels to the target channel scales exponentially with the number of noisy gates: $\sigma \propto \gamma^{N_{\text{noisy}}}$, cf.~Eq.~\eqref{eq:prefactor_NT_NEC}.

\subsubsection{PEC and PER as particular cases of NT}
\label{sec:PEC}
\par Probabilistic Error Cancellation (PEC) corresponds to the limiting case of NT where the noiseless channel is chosen as target: $E^\text{target}_2=\mathbb{I}_2$. Therefore for all Pauli directions $a$, $f^\text{PEC}_a=(f^\text{gate}_a)^{-1}>1$, which necessarily leads to quasiprobability distributions in the $\chi$-matrix representation, according to Eq.~\eqref{eq:WH}. Consequently, the noisier the quantum gate, the greater $\gamma_{\text{PEC}}$; with the error rates of entangling gates in current NISQ devices, PEC quickly leads to intractable sampling costs for circuits with many noisy $2$-qubit gates.

In an effort to alleviate this sampling cost, it has been proposed to only partially reduce or even increase the noise strength to various levels using Probabilistic Error Reduction (PER), and then perform ZNE to mitigate the effect of the remaining noise by extrapolating results to the zero-noise limit. While ZNE usually exhibits better scaling than other error mitigation protocols, it may fail to perfectly mitigate the noise effects, because of the approximations necessary to obtain a reasonable extrapolation ansatz function. Nevertheless, combining both methods has led to remarkable achievements in the efficiency of error mitigation, for instance in Ref.~\cite{kim2023}, which partially motivated the use of NT with another error mitigation scheme, NEC, in the present work.

\subsubsection{Noise-tailoring towards a depolarizing noise channel}
\label{app:DNT}

The formalism developed above can be used, for example, to drive an initial Pauli noise channel towards a depolarizing noise channel, a particular case of isotropic Pauli noise channel. This is what we did in the present work to enhance the potentiality of NEC, a mitigation technique particularly efficient in mitigating this type of noise. We take as the target channel a $2$-qubit depolarizing noise channel, which can be written as:
\begin{eqnarray}
    \mathcal{E}_2^\text{depo.}(\rho)&=&(1-p)\rho+\frac{p}{15}\sum_{a=1}^6 P^{(2)}_a\,\rho\,P^{(2)}_a\nonumber\\
    &=&(1-\epsilon)\rho+\epsilon\frac{\mathbb{I}_{4}}{4}
    \label{eq:depo_noise2}
\end{eqnarray} where $\epsilon=16p/15$, and the $\sigma_i$ are the Pauli matrices ($\sigma_0=\mathbb{I}_1$). Its PTM representation is a $16\times 16$ diagonal matrix:
\begin{equation}
  E_2^\text{target}\equiv E_2^\text{depo.}(\epsilon)=\text{diag}(1,1-\epsilon,\cdots,1-\epsilon).
\end{equation}
After the RC average, the application of a CNOT gate induces a Pauli noise channel with the following PTM representation:
\begin{equation}
    E_2^\text{gate}=\text{diag}(1,f_1^\text{gate},\cdots,f_{16}^\text{gate}).
\end{equation}
Thus, using Eq.~\eqref{eq:noise_target}, the fidelities of the stochastic noise channel to implement are:
\begin{equation}
   f_a^\text{target}(\epsilon)=\frac{1-\epsilon}{f_a^\text{gate}}
\end{equation}
for $a\neq0$, and $f_0^\text{target}=1$. Employing the Walsh-Hadamard transformation, Eq.~\eqref{eq:WH}, yields quasi-probabilities $q_a^\text{DNT}(\epsilon)$:
\begin{equation}
q_a^{\mathrm{DNT}}(\epsilon)
= \frac{1}{16}\!\left[
1 \;+\; (1-\epsilon)\!\sum_{b=1}^{15}
(-1)^{\langle P_a^{(2)}, P_b^{(2)} \rangle_{\mathrm{sp}}}
\,\frac{1}{f_b^{\mathrm{gate}}}
\right],
\label{eq:q_DNT}
\end{equation}
with sampling factor
\begin{equation}
\gamma^{\mathrm{DNT}}(\epsilon)
= 1 + 2\sum_{q_a^{\mathrm{DNT}}(\epsilon)<0}\bigl|q_a^{\mathrm{DNT}}(\epsilon)\bigr|.
\label{eq:gamma_DNT}
\end{equation}

The sampling cost $\gamma^{\mathrm{DNT}}(\epsilon)$ in Eq.~\eqref{eq:gamma_DNT} is fully determined by the contribution of the negative quasi-probabilities in Eq.~\eqref{eq:q_DNT}. For instance, a practical way to minimize the Monte Carlo overhead (aside from the case of increasing noise magnitude) is to match the magnitude of the target depolarizing PTM to the original Pauli channel by choosing
\begin{align}
\epsilon
= 1-\overline{f}^{\,\mathrm{gate}},
\label{eq:target_depo_ideal}
\end{align}
$\overline{f}^{\,\mathrm{gate}}$ being the average of the original gate fidelities. This “matched” depolarizing target typically yields quasi-probabilities that are closer to non-negative than in PEC ($\epsilon=0$) or PER (where typically $f_b^{\,\mathrm{target}}
> \overline{f}^{\,\mathrm{gate}}$ for all $b$), thereby reducing the number and magnitude of negative coefficients and hence the sampling cost. In the main text, the error rates of the target depolarizing noise channels for each junction are chosen in order to minimize the sampling cost of the overall error mitigation protocol. This is typically achieved by choosing $\epsilon
< 1-\overline{f}^{\,\mathrm{gate}}$, as explained in Sec.~\ref{sec:NEC}.

\subsection{NT accounting for crosstalk (cNT)}
\label{app:crosstalk}

Superconducting quantum chips are known to be plagued by unwanted resonances occurring with neighboring qubits during the application of entangling gates, which induces coherent errors and strongly affects the system's overall coherence. This effect is known as (spillover) crosstalk and strongly limits the performances of such devices \cite{Sarovar2020,Ding2020,Murali2020,kandala2021,Rudinger2021,Xie2021,Zhao2022,ketterer2023}. To mitigate this issue, we introduced in a previous work~\cite{perrin2024} the concept of crosstalk randomized compiling (cRC), which allows for the mapping of noise on neighboring qubits into a depolarizing channel. This is essentially achieved by adding additional random $\frac{\pi}{2}$ rotation gates to the RC gate set. Using this technique, coherent noise on both the active and neighboring qubits of a CNOT gates can be effectively turned into a mixture (see below) of depolarizing (on the neighboring qubit) and Pauli noise channels (on the active qubit), on top of which NT can be applied.

We now extend the NT procedure from Appendix~\ref{app:DNT} to the case where crosstalk is present, matching the actual protocol implemented in the main text. Considering that crosstalk requires extending the noise channel for each CNOT gates to its neighboring qubits, i.e., to set $q=4$ for a linear layout. Since the benchmark experiments presented in this work are limited to $3$ qubits, we consider here, without loss of generality, the case where CNOT gates have only one neighbor, and set $q=3$.

After applying cRC, the $3$-qubit noise channel for the CNOT gate and its neighboring qubit is a mixture of a Pauli noise channel on the active qubits of the CNOT and a depolarizing noise channel on the neighboring qubit that has $31$ free parameters~\cite{perrin2024}. Indeed, if we again denote as $\{P_a^{(2)}\}_{a=0}^{15}$ the $2$-qubit Pauli string basis on the CNOT qubits (with $a=0$ corresponding to $I\!\otimes I$), and let the neighboring index be $\kappa\in\{I,X,Y,Z\}$, the fidelities of the total noise channels can be indexed as $f_{a\kappa}$. In this basis the $3$–qubit PTM is diagonal and, since the cRC leaves the neighboring noise channel isotropic, it depends only on whether $\kappa=I$ or $\kappa\in\{X,Y,Z\}$ (by definition for a single-qubit noise channel). Writing two $16$–dimensional vectors of fidelities as

\begin{align}
\mathbf F^{\mathbb I}=\bigl[ f_{a{\mathbb I}}\bigr]_{a=0}^{15}\quad\text{and}\quad
\mathbf F^{\mathrm D}=\bigl[f_{a{\mathrm D}}\bigr]_{a=0}^{15},
\end{align}
with $f_{0{\mathbb I}}=1$ required for trace preservation of the system's density matrix, and $\mathrm D$ denoting the case of a depolarizing error of any type on the neighboring qubit, we have

\begin{align}
E_3^{\mathrm{gate}}
=\mathrm{diag}\!\bigl(\mathbf F^{\mathbb I},\,\mathbf F^{\mathrm D},\,\mathbf F^{\mathrm D},\,\mathbf F^{\mathrm D}\bigr),
\label{eq:E3gate_blocks}
\end{align}
where $\mathrm{diag}(\cdot)$ concatenates the four length–16 blocks along the diagonal in the order set by $\kappa=I,X,Y,Z$. The diagonal of $E_3^{\mathrm{gate}}$ is organized into four \(16\)-entry blocks indexed by the neighboring depolarizing channel, so the three non–identity blocks are identical. Hence since there are $15$ free parameters in $\mathbf F^{\mathbb I}$ (all $a\neq 0$) and $16$ in $\mathbf F^{\mathrm D}$, totaling $31$ free parameters.

In the NT procedure used in the main text, we use as a target noise channel the following $3$-qubit quasi-local depolarizing noise channel, originally introduced in \cite{perrin2024} as an approximation of the outcome of cRC:

\begin{align}
\mathcal{E}_3(\rho)&=(1-\epsilon_{\text{CNOT}}-\epsilon_{\text{neigh.}}-\epsilon_{\text{glob.}})\rho \nonumber\\
&+\epsilon_{\text{CNOT}}\frac{\mathbb{I}_{01}}{4}\otimes\text{Tr}_{01}(\rho)\nonumber\\
&+\epsilon_{\text{neigh.}}\text{Tr}_{2}(\rho)\otimes\frac{\mathbb{I}_{2}}{2}+\epsilon_{\text{glob.}}\frac{\mathbb{I}_{012}}{8},
\label{eq:3qubitchannel}
\end{align}
where $0$ and $1$ denote the indices of the active qubits of the CNOT gate, and $2$ is the index of the neighboring qubit, $\text{Tr}_{\{k\}}$ is the partial trace over the set of qubits $\{k\}$. $\epsilon_{\text{CNOT}}$ (resp. $\epsilon_{\text{neigh.}}$ and $\epsilon_{\text{glob.}})$ denotes the active qubits' (resp. neighboring qubit and global) depolarizing error rate. In the PTM representation, it reads:

\begin{align}
  E_3^\text{depo.}=\text{diag}(\textbf{D}^0,\textbf{D}^1,\textbf{D}^1,\textbf{D}^1)
\end{align}
with
\begin{align}
\textbf{D}^0=
   \begin{pmatrix}
       1\\
       1-\epsilon_\text{CNOT}-\epsilon_\text{glob.}\\
       \vdots\\
       1-\epsilon_\text{CNOT}-\epsilon_\text{glob.}
   \end{pmatrix}\nonumber\\\nonumber\\
\end{align}
and
\begin{align}
\mathbf{D}^1=
   \begin{pmatrix}
       1-\epsilon_\text{neigh.}-\epsilon_\text{glob.}\\
   1-\epsilon_\text{neigh.}-\epsilon_\text{CNOT}-\epsilon_\text{glob.}\\
   \vdots\\
   1-\epsilon_\text{neigh.}-\epsilon_\text{CNOT}-\epsilon_\text{glob.}\nonumber
   \end{pmatrix}.
\end{align}

This noise channel has only three parameters, and is much friendlier to error mitigation than its Pauli counterpart described above. As discussed in Sec.~\ref{sec:NEC}, the three depolarizing parameters are chosen to minimize the overall mitigation cost.

Using the PTM representation of the post-cRC and the target noise channels above, the NT procedure follows the same steps as the case without crosstalk from the last subsection: using Eq.~\eqref{eq:noise_target}, we obtain the fidelities of the stochastic noise channel  $E_3^\text{DNT}$. The $3$-qubit Walsh-Hadamard transformation maps these fidelities to quasi-probabilities of Pauli errors needed to be stochastically implemented after each CNOT gate, as well as the neighboring qubit. This process can be extended to the $4$-qubit case, and even to an arbitrary number of neighboring qubits, potentially encompassing high-connectivity situations and/or higher-order crosstalk effects in a straightforward way.

\section{Pauli noise tomography}
\label{app:PNT}

A crucial aspect for the implementation of the NT protocol is a precise characterization of the noise. In this section we describe our method to efficiently perform the tomography of the Pauli noise channel affecting a CNOT gate in real quantum devices. While Pauli Noise Tomography (PNT) is acknowledged and discussed in the literature ~\cite{mcdonough2022,vandenberg2023,Chen2023}, the explicit circuit constructions required for the practical implementation of the tomography, particularly in the presence of crosstalk, are not readily available. In this Appendix, we provide here our version of these circuits along with the mathematical description of PNT. Once again, the derivation is done for the CNOT gate, but can be extended to any type of $2$-qubit gate.

\subsection{Without crosstalk}
\label{app:PNT_no_crosstalk}

In the absence of crosstalk, the aim of PNT is to fully characterize the Pauli noise channels affecting the target qubits of the CNOT gate. The goal is to estimate the value the $15$ independent fidelities of the effective Pauli noise channel (obtained after RC) affecting the CNOT gates of a given physical qubit junction from the smallest possible number of circuit experiments.

\begin{figure}[h!]
    \centering
\includegraphics[scale=0.5]{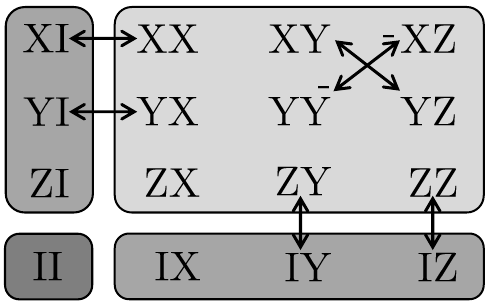}
    \caption{Pauli transfer diagram of the CNOT gate. The four Pauli strings in the bottom left corner that are not linked by an arrow are left invariant by the CNOT gate. The minus sign on the arrow joining the YY and XZ directions indicates the phase $\pi$ induced by the CNOT gate action on these states. Figure inspired by Ref.~\cite{vandenberg2023}.}
    \label{fig:diagram}
\end{figure}

Let us start by reminding a few useful properties of the CNOT gate. First of all, the CNOT gate squares to identity: $\text{CNOT}^2=\mathbb{I}_2$. In particular, this means that a sequence of $2n+1$ CNOT gates with integer $n$ remains logically equivalent to an isolated CNOT gate, but with amplified noise levels when the gate is noisy. This property has been widely used in the context of ZNE as a simple method to amplify the noise affecting a given circuit. Second, the CNOT gate is a Clifford gate, therefore it maps a Pauli string to a Pauli string. Its \textit{Pauli transfer diagram} that encodes this mapping is shown in Fig.~\ref{fig:diagram} (notice the minus sign accompanying the transformation between YY and XZ).

Pauli strings both form a basis for the system's density matrix and serve to define the PTM of noise, cf.~App.~\ref{app:noise_representation}. Imagine preparing the $2$ qubits in a state with density matrix equal to a certain Pauli string, $\rho = P_a$ (which is evidently not physical, but useful for the below consideration). The application of an ideal CNOT gate will yield another Pauli string $P_b$ according to the diagram in Fig.~\ref{fig:diagram}. A non-ideal CNOT gate will yield $f_b P_b$ instead, where $f_b$ is the respective fidelity of the noise channel.

Based on this observation, we build in the following a number of circuits required to extract all $15$ parameters. For the sake of clarity, we assume that the qubits are initially prepared in the $\ket{00}$ product states, and that the end-of-circuit measurements are performed in the ZZ basis, as by default in IBMQ devices. Also, we assume that the CNOT gate is applied here with the first (top) qubit as the control qubit and the second (bottom) qubit as the active qubit, as depicted in the following quantum circuit plots. For completeness, the procedure should also be performed with reversed CNOT gates, particularly in the crosstalk case, where crosstalk fidelities may depend strongly on the direction of the gate.

\begin{figure}[h!]
    \centering
    \includegraphics[scale=0.2]{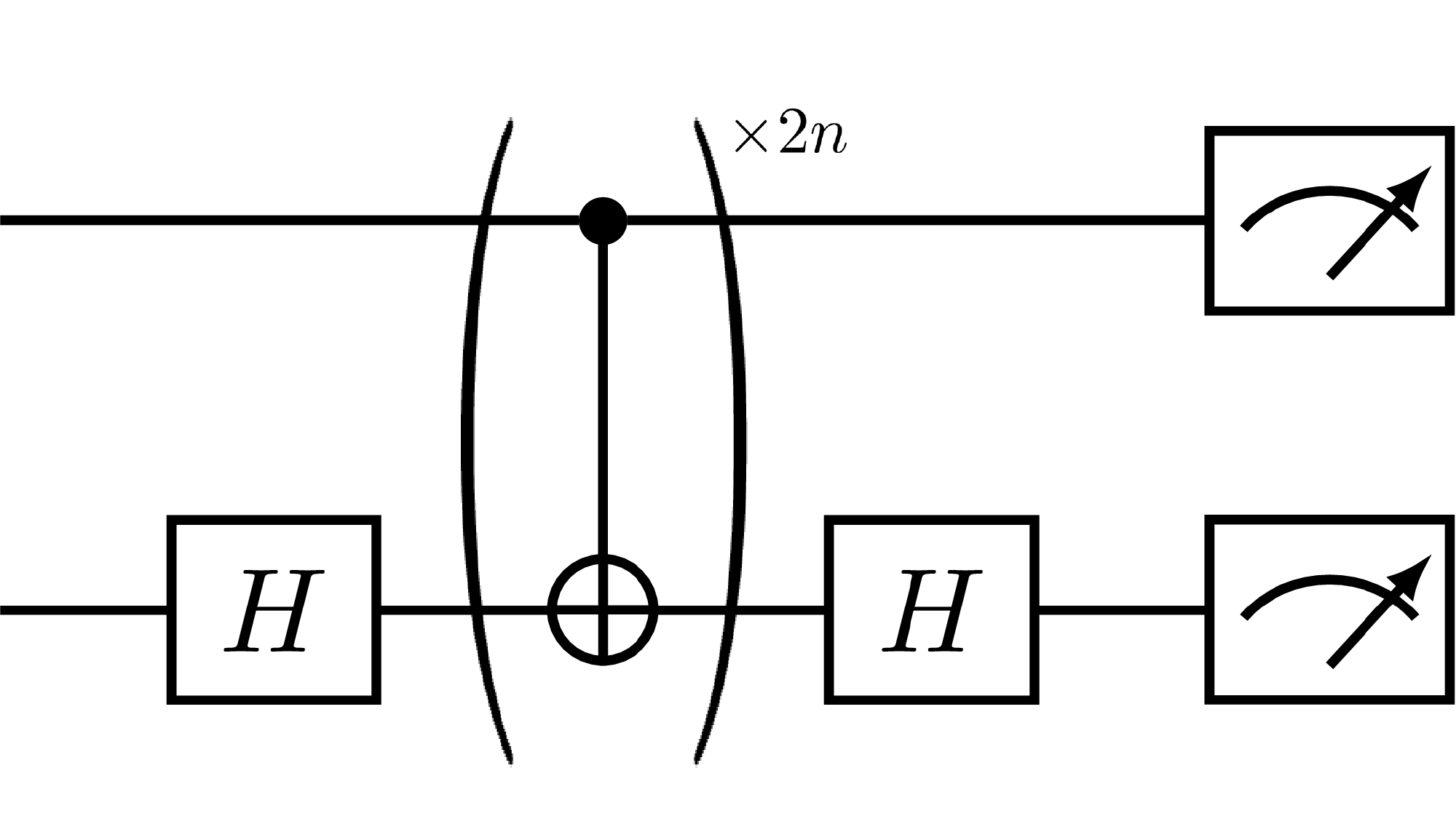}
    \caption{Circuit preparing the state $\ket{0+}$ and measuring the control qubit in the $z$ direction and the target qubit in the X direction. It gives access to the fidelities $(f_{\text{IX}})^{2n}$, $(f_{\text{ZI}})^{2n}$ and $(f_{\text{ZX}})^{2n}$. The measurement symbol at the end of the circuit depicts a measurement of the qubit in the Z-basis.}
    \label{fig:inv_fid}
\end{figure}

We start by the simplest case, estimating the fidelities along the directions $\text{II}$, $\text{IX}$, $\text{ZI}$, and $\text{ZX}$ which are left invariant by the CNOT gate. Indeed, by applying a Hadamard gate on the second qubit, we can easily prepare the state $\ket{0+}$, with density matrix:
\begin{align}
\rho = \ket{0 +}\bra{0+}=\frac{1}{4}(\text{II}+\text{IX}+\text{ZI}+\text{ZX}).
\end{align}
After the application of a noisy CNOT gate with Pauli noise channel $E^{\text{gate}}$, the state becomes:
\begin{align}
\frac{1}{4}(\text{II}+f_{\text{IX}}\text{IX}+f_{\text{ZI}}\text{ZI}+f_{\text{ZX}}\text{ZX}),
\end{align}
where we have unraveled the subscript $a$ into its two components on the first and second qubit (IX, ZI, ZX). Then, measuring simultaneously the first qubit in the Z direction and the second qubit in the X direction yields the quantity $f_{\text{ZX}}$. Likewise, solely measuring the first (resp. second) qubit in the Z (resp.~X) direction provides the value of $f_{\text{ZI}}$ (resp.~$f_{\text{IX}}$). To improve the estimation accuracy, one can repeat the CNOT gate $n$ times (with $n>1$), so that the measured expectation values scale as $(f_a)^n$. In this case, small deviations of $f_a$ from unity are amplified as $1 - (f_a)^n \simeq n(1 - f_a)$, allowing the extraction of $f_a$ with a precision that improves approximately linearly with $n$ for $f_a \approx 1$. A simple exponential fit of the measured signal versus $n$ then provides an accurate estimate of the corresponding fidelities.
Following this idea yields the quantum circuit depicted in Fig.~\ref{fig:inv_fid} for the characterization of the fidelities $(f_{\text{IX}})$, $(f_{\text{ZI}})$ and $(f_{\text{ZX}})$ (in the Figure, an amplification factor of $2n$ instead of $n$ is used, for consistency with other circuits described below).

For the Pauli directions that are non-invariant terms under the action of the CNOT gate, things are slightly more complicated. They fall into two categories . The first one contains basis states XY, YZ, XZ and YY (top-right four in Fig.~\ref{fig:diagram}). The remaining Pauli basis states XX, XI, YX, YI, ZY, IY, ZZ, and IZ belong to the second category.

For the first category, corresponding fidelities can be uniquely measured. The trick is to apply two single-qubit $\frac{\pi}{2}$-rotation gates after the application of the CNOT gate to go back to the original state. For instance, applying the CNOT gate to an XY density matrix term yields the state YZ, but the original XY state can be restored by applying two single-qubit $\frac{\pi}{2}$-rotation gates.
However, implementing a density state proportional to XY is not possible (if we impose the state preparation to be a pure state). Instead, starting from the state $\ket{++_y}$ (where $+_y$ refers to the eigenvector of $\sigma_y$) and applying the appropriate rotations, we can obtain the desired state:

\begin{align}
&\ket{++_y}\bra{++_y}=\frac{1}{4}(\text{II}+\text{XI}+\text{IY}+\text{XY})\\
&\xrightarrow{\text{CNOT}}\frac{1}{4}(\text{II}+f_{\text{XX}}\text{XX}+f_{\text{ZY}}\text{ZY}+f_{\text{YZ}}\text{YZ})\nonumber\\
&\xrightarrow{R_z\left(\frac{\pi}{2}\right)\otimes R_x\left(\frac{\pi}{2}\right)}\frac{1}{4}(\text{II}-f_{\text{XX}}\text{YX}-f_{\text{ZI}}\text{ZX}+f_{\text{YZ}}\text{XY})\nonumber
\end{align}

\begin{figure*}[h!]
    \centering
\begin{tabular}{@{}c@{\hspace{1.2cm}}c@{}}
  \includegraphics[height=2.8cm,trim={0 3cm 0 3cm},clip]{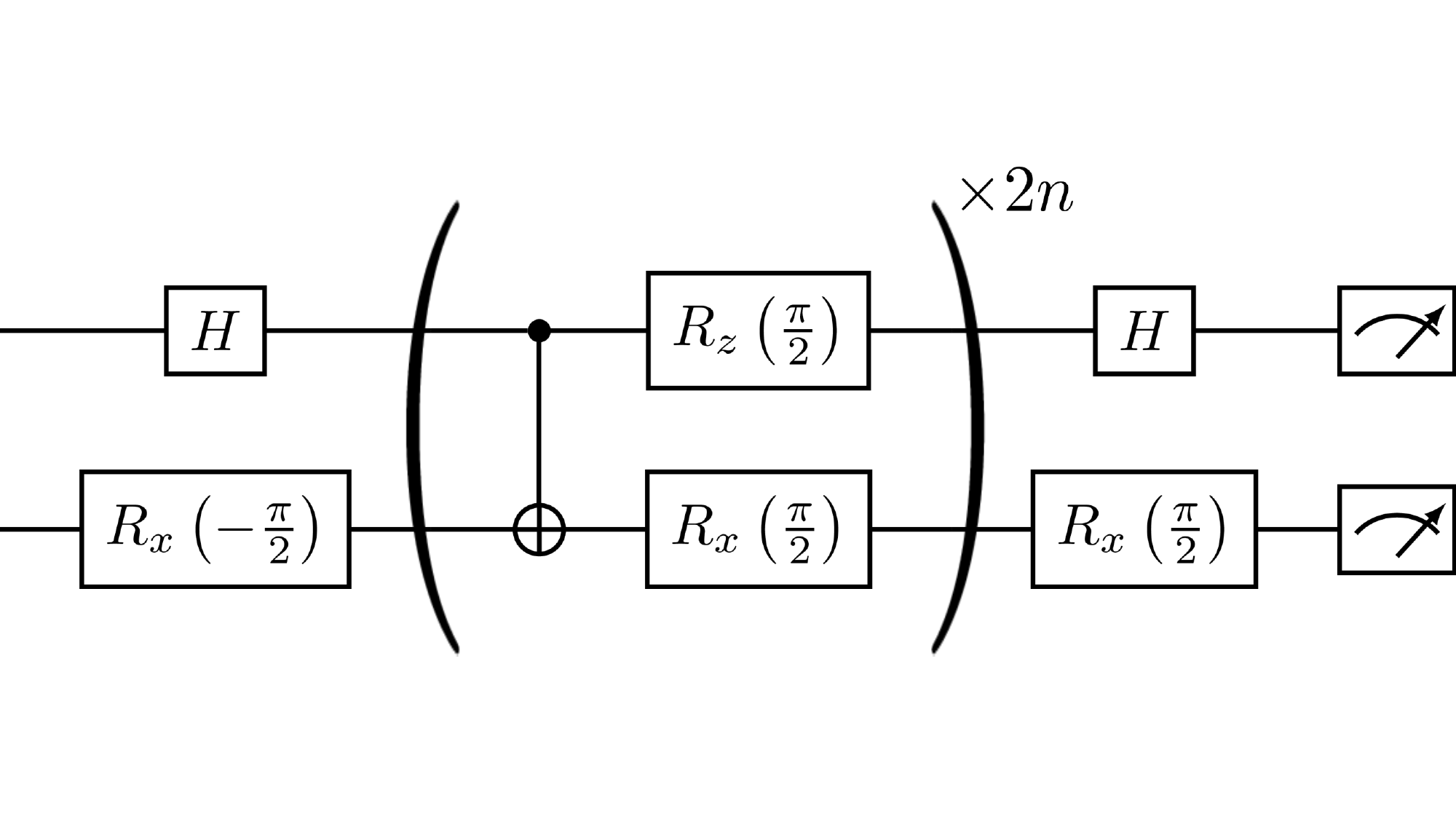}   & \includegraphics[height=2.8cm,trim={0 3cm 0 3cm},clip]{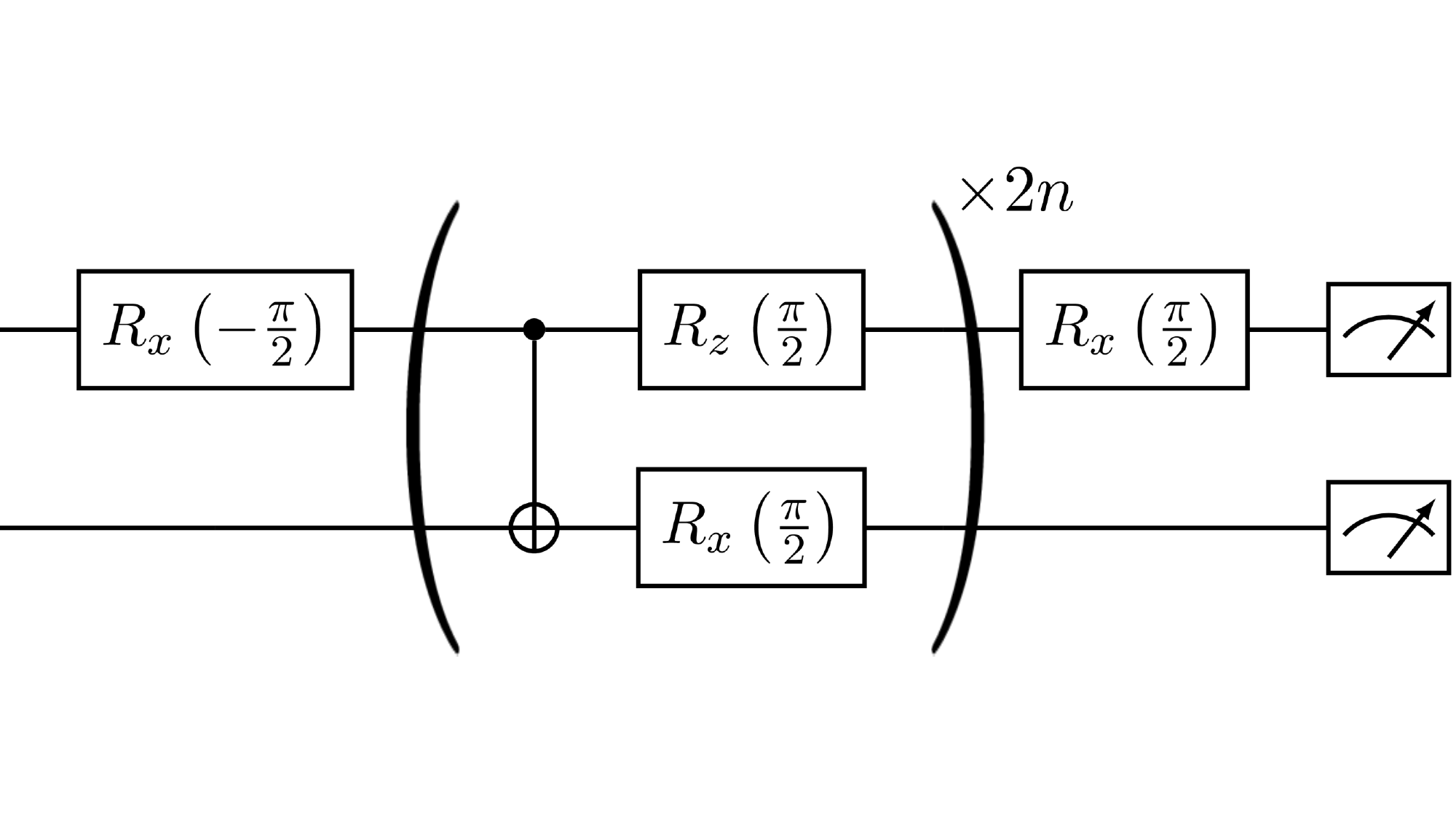} \\
  \includegraphics[height=2.8cm,trim={0 3cm 0 3cm},clip]{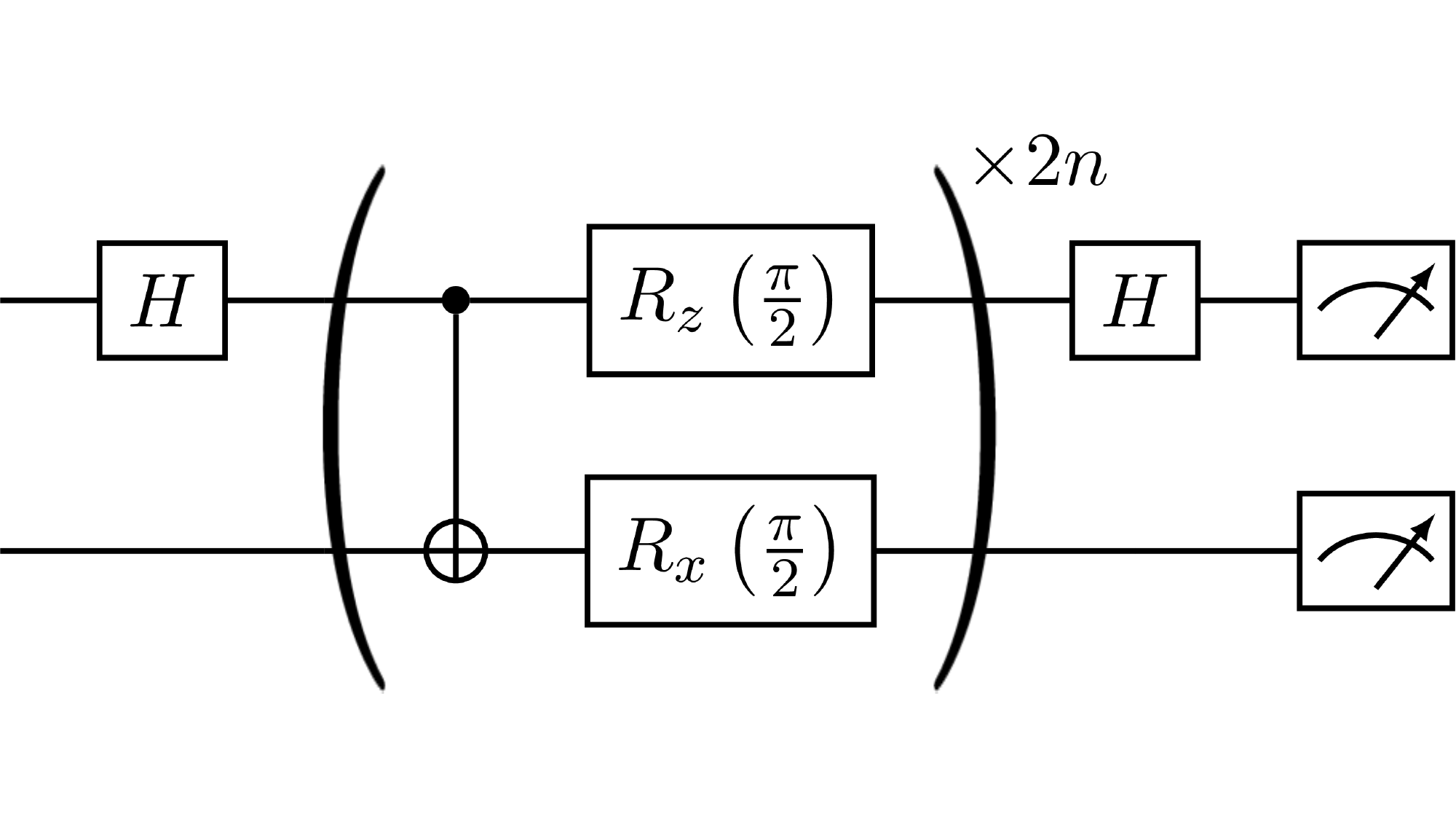}   & \includegraphics[height=2.8cm,trim={0 3cm 0 3cm},clip]{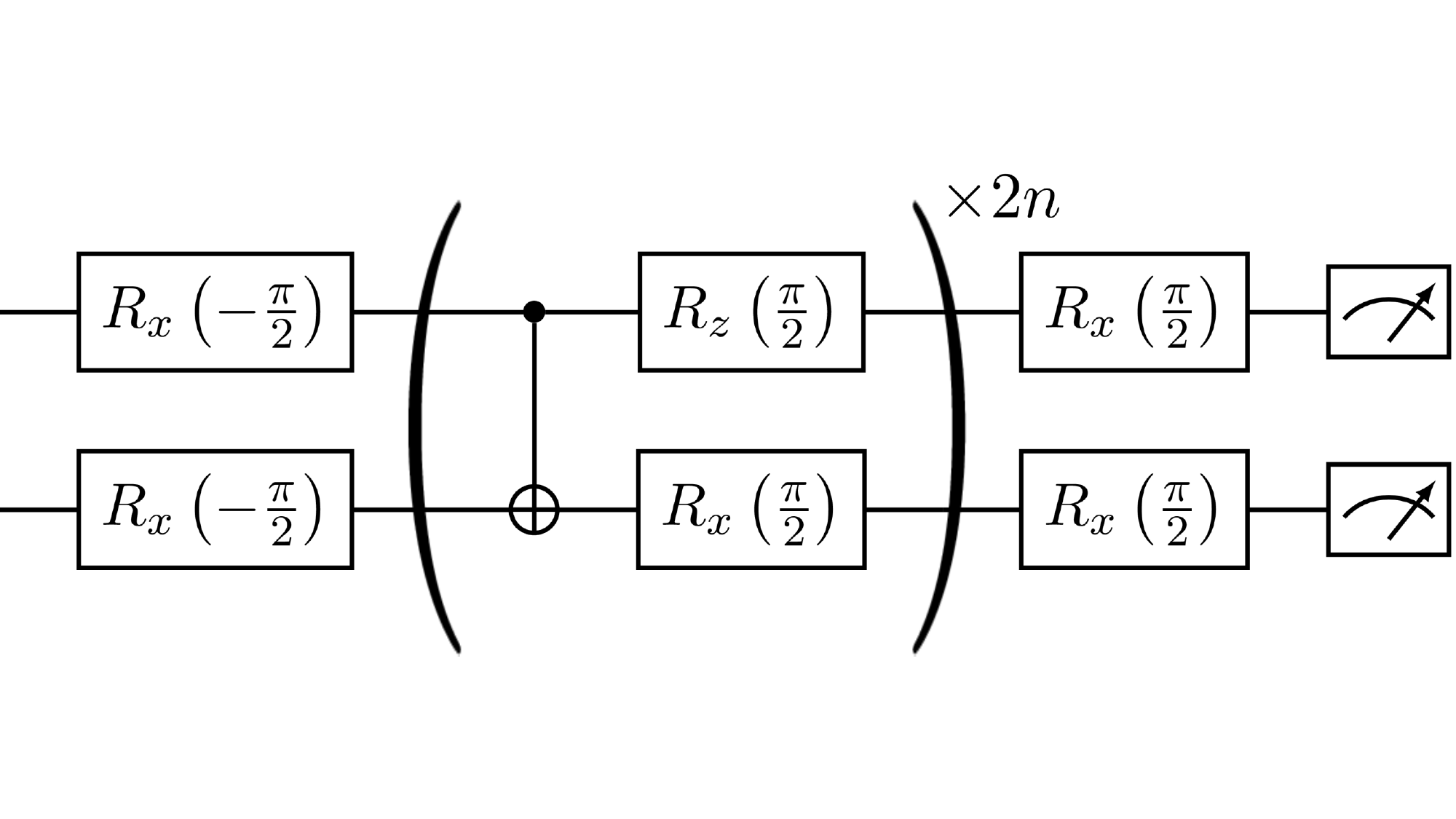} \\
\end{tabular}
    \caption{ Circuits preparing states $\ket{++_y}$ (top left), $\ket{+_y0}$ (top right),  $\ket{+0}$ (bottom left) or $\ket{+_y+_y}$ (bottom right). States are measured in the same direction as they are prepared. One deduces fidelities $f_{\text{YZ}}^{2n}$ (top left),  $f_{\text{XY}}^{2n} $ (top right),  $f_{\text{YY}}^{2n}$ (bottom left) and $f_{\text{XZ}}^{2n}$ (bottom right).}
    \label{fig:cat1}
\end{figure*}

Thus, measuring the first qubit in the X direction and the second qubit in the Y direction indeed yields the fidelity $f_{\text{YZ}}$. Similarly to the previous case, applying this procedure repeatedly leads to $f_{\text{YZ}}^n$ where $n$ is the number of CNOT gates applied (see the top left circuit of Fig.~\ref{fig:cat1}), and yields a precise estimate of the fidelity from an exponential fit against increasing values of $n$. In the same manner, we obtain the fidelities $f_{\text{XY}}$, $f_{\text{YY}}$ and $f_{\text{XZ}}$ using the following procedures:

\begin{align}
&\ket{+_y0}\bra{+_y0}=\frac{1}{4}(\text{II}+\text{YI}+\text{IZ}+\text{YZ})\\
&\xrightarrow{\text{CNOT}}\frac{1}{4}(\text{II}+f_{\text{YX}}\text{YX}+f_{\text{ZZ}}\text{ZZ}+f_{\text{XY}}\text{XY})\nonumber\\
&\xrightarrow{R_z\left(\frac{\pi}{2}\right)\otimes R_x\left(\frac{\pi}{2}\right)}\frac{1}{4}(\text{II}-f_{\text{YX}}\text{XX}-f_{\text{ZZ}}\text{ZY}+f_{\text{XY}}\text{YZ})\nonumber
\end{align}

\begin{align}
&\ket{+0}\bra{+0}=\frac{1}{4}(\text{II}+\text{XI}+\text{IZ}+\text{XZ})\\
&\xrightarrow{\text{CNOT}}\frac{1}{4}(\text{II}+f_{\text{XX}}\text{XX}+f_{\text{ZZ}}\text{ZZ}-f_{\text{YY}}\text{YY})\nonumber\\
&\xrightarrow{R_z\left(\frac{\pi}{2}\right)\otimes R_x\left(\frac{\pi}{2}\right)}\frac{1}{4}(\text{II}+f_{\text{XX}}\text{YX}-f_{\text{ZZ}}\text{ZY}+f_{\text{YY}}\text{XZ})\nonumber
\end{align}

\begin{align}
&\ket{+_y+_y}\bra{+_y+_y}=\frac{1}{4}(\text{II}+\text{YI}+\text{IY}+\text{YY})\\
&\xrightarrow{\text{CNOT}}\frac{1}{4}(\text{II}+f_{\text{YX}}\text{YX}+f_{\text{ZY}}\text{ZY}-f_{\text{XZ}}\text{XZ})\nonumber\\
&\xrightarrow{R_z\left(\frac{\pi}{2}\right)\otimes R_x\left(\frac{\pi}{2}\right)}\frac{1}{4}(\text{II}-f_{\text{YX}}\text{XX}+f_{\text{ZY}}\text{ZZ}+f_{\text{XZ}}\text{YY}).\nonumber
\end{align}
The respective circuits are shown in Fig.~\ref{fig:cat1}.

For the second category, obtaining individual fidelities is not as straightforward. Note that applying the CNOT gate twice restores the original Pauli string state. Consequently, applying an even number of CNOT gates, pairs of fidelities can be measured:
$(f_{\text{XX}}f_{\text{XI}})^n$, $(f_{\text{YX}}f_{\text{YI}})^n$, $(f_{\text{ZY}}f_{\text{IY}})^n$ and $(f_{\text{ZZ}}f_{\text{IZ}})^n$. However, a degeneracy on the exact value of each individual fidelity remains.  Measuring an odd number of CNOT gates solves this issue by giving access to the following quantities:
$(f_{\text{XX}}f_{\text{XI}})^n f_{\text{XX}}$, $(f_{\text{XX}}f_{\text{XI}})^n f_{\text{XI}}$, $(f_{\text{YX}}f_{\text{YI}})^nf_{\text{YI}}$, $(f_{\text{YX}}f_{\text{YI}})^nf_{\text{YX}}$,$(f_{\text{ZY}}f_{\text{IY}})^n f_{\text{IY}}$, $(f_{\text{ZY}}f_{\text{IY}})^n f_{\text{ZY}}$,$(f_{\text{ZZ}}f_{\text{IZ}})^nf_{\text{IZ}}$ and $(f_{\text{ZZ}}f_{\text{IZ}})^nf_{\text{ZZ}}$. For instance, let us start from the state $\ket{++}$ following the application of two single qubit gates. Then, applying an odd number of CNOT gates and measuring both qubits in the X direction leads to the quantity $(f_{\text{XX}}f_{\text{XI}})^n f_{\text{XX}}$. If only the measurement on the first qubit is taken into account, one obtains $(f_{\text{XX}}f_{\text{XI}})^n f_{\text{XI}}$. Indeed:

\begin{align}
  &\ket{++}\bra{++}=\frac{1}{4}(\text{II}+\text{XI}+\text{IX}+\text{XX})\label{eq:XX}\\
  &\qquad\qquad\qquad\qquad\qquad\xdownarrowright[\text{CNOT}^{2n+1}]{2em}\nonumber\\
  &\frac{1}{4}\!\left(\text{II}+(f_{\text{XX}}f_{\text{XI}})^n f_{\text{XX}}\text{XX}
  +f_{\text{IX}}^{2n+1}\text{IX}+(f_{\text{XX}}f_{\text{XI}})^n f_{\text{XI}}\text{XI}\right)\nonumber
\end{align}

These measurements uniquely fix the value of $f_{\text{XI}}$ and $f_{\text{XX}}$. Similarly, one can deduce the other remaining fidelities:

\begin{align}
    &\ket{+_y+}\bra{+_y+}=\frac{1}{4}(\text{II}+\text{YI}+\text{IX}+\text{YX})\label{eq:YX}\\
    &\qquad\qquad\qquad\qquad\qquad\xdownarrowright[\text{CNOT}^{2n+1}]{2em}\nonumber\\
    &\frac{1}{4}(\text{II}+(f_{\text{YX}}f_{\text{YI}})^n f_{\text{YX}}\text{YX}+f_{\text{IX}}^{2n+1}\text{IX}+(f_{\text{YX}}f_{\text{YI}})^n  f_{\text{YI}}\text{YI})\nonumber
\end{align}

\begin{align}
    &\ket{0+_y}\bra{0+_y}=\frac{1}{4}(\text{II}+\text{ZI}+\text{IY}+\text{ZY})\label{eq:ZY}\\
    &\qquad\qquad\qquad\qquad\qquad\xdownarrowright[\text{CNOT}^{2n+1}]{2em}\nonumber\\
    &\frac{1}{4}(\text{II}+f_{\text{ZI}}^{2n+1}\text{ZI}+(f_{\text{IY}}f_{\text{ZY}})^n f_{\text{ZY}}\text{ZY}+(f_{\text{IY}}f_{\text{ZY}})^n f_{\text{IY}}\text{IY})\nonumber
\end{align}

\begin{align}
    &\ket{00}\bra{00}=\frac{1}{4}(\text{II}+\text{ZI}+\text{IZ}+\text{ZZ})\label{eq:ZZ}\\
    &\qquad\qquad\qquad\qquad\qquad\xdownarrowright[\text{CNOT}^{2n+1}]{2em}\nonumber\\
    &\frac{1}{4}(\text{II}+f_{\text{ZI}}^{2n+1}\text{ZI}+(f_{\text{ZZ}}f_{\text{IZ}})^n f_{\text{ZZ}}\text{ZZ}+(f_{\text{ZZ}}f_{\text{IZ}})^n f_{\text{IZ}}\text{IZ})\nonumber
\end{align}

Note that by solely considering the measurement of the first qubit in Eqs.~\eqref{eq:XX} and~\eqref{eq:YX} or the second qubit in Eqs.~\eqref{eq:ZY} and~\eqref{eq:ZZ}, one obtains two extra estimations of the fidelities $f_{\text{ZI}}$ and $f_{\text{IX}}$.

In conclusion, employing only nine (repeated with different $n$) circuits, we have measured nineteen observables, which allow to extract all individual fidelities by taking into account four redundancies. Running the circuits at various depth $n$, and performing a global fitting procedure we indeed deduce the fifteen different fidelities of the 2-qubit Pauli noise channel.

\begin{figure*}[h!]
    \centering
\begin{tabular}{@{}c@{\hspace{1.2cm}}c@{}}
  \includegraphics[height=2.8cm,trim={0 2cm 0 2cm},clip]{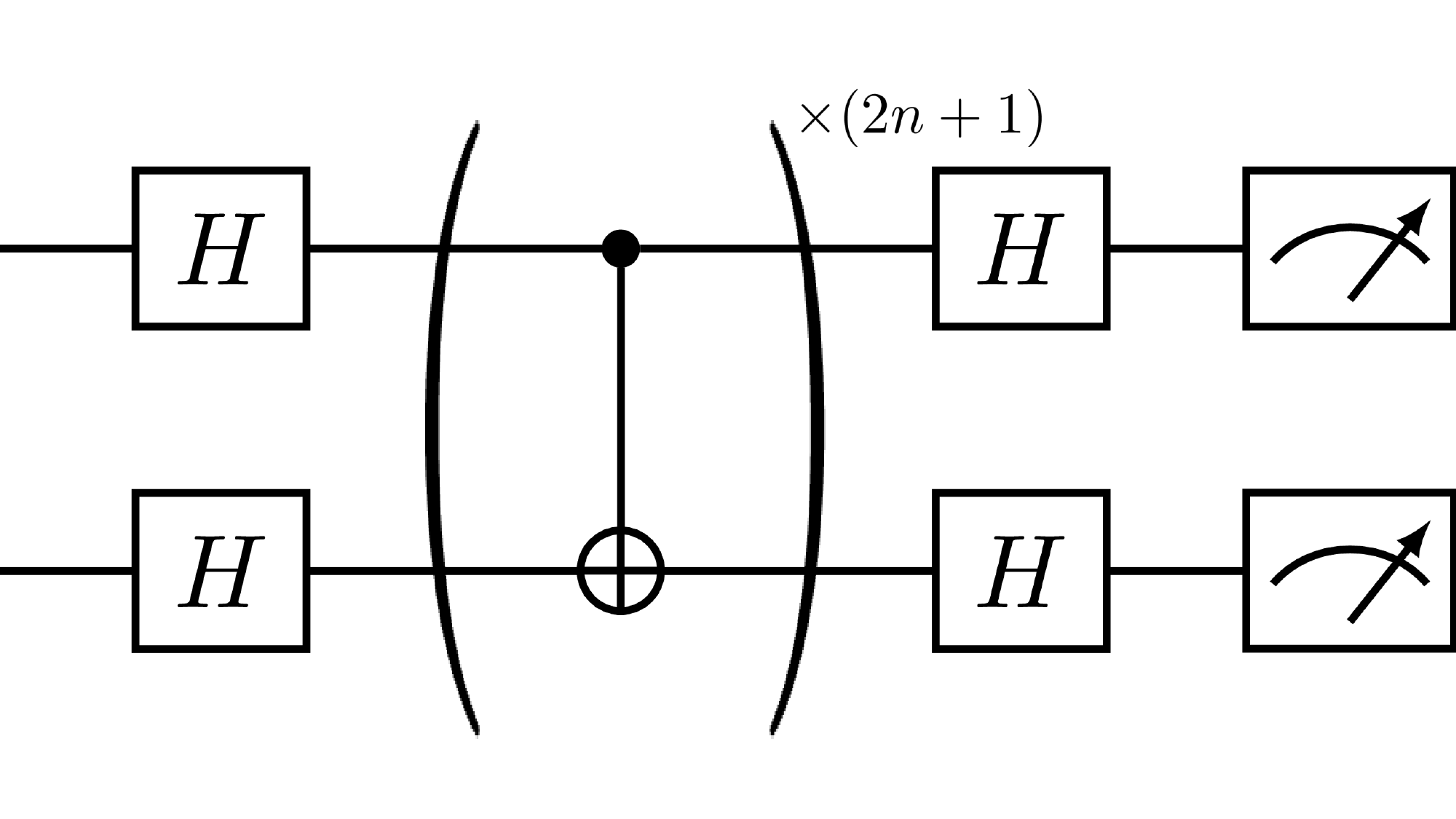}   & \includegraphics[height=2.8cm,trim={0 3cm 0 3cm},clip]{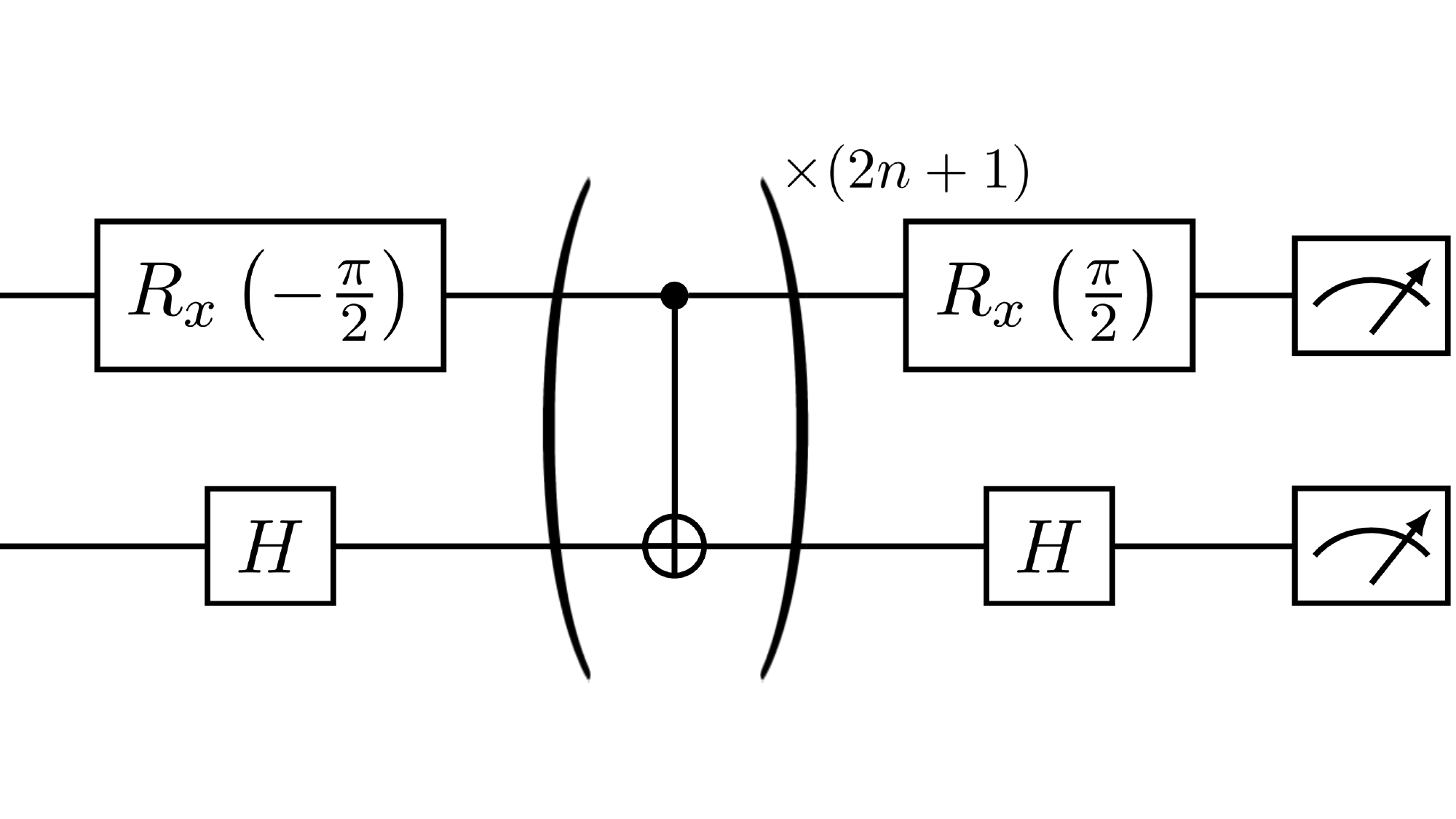} \\
  \includegraphics[height=2.8cm,trim={0 3cm 0 3cm},clip]{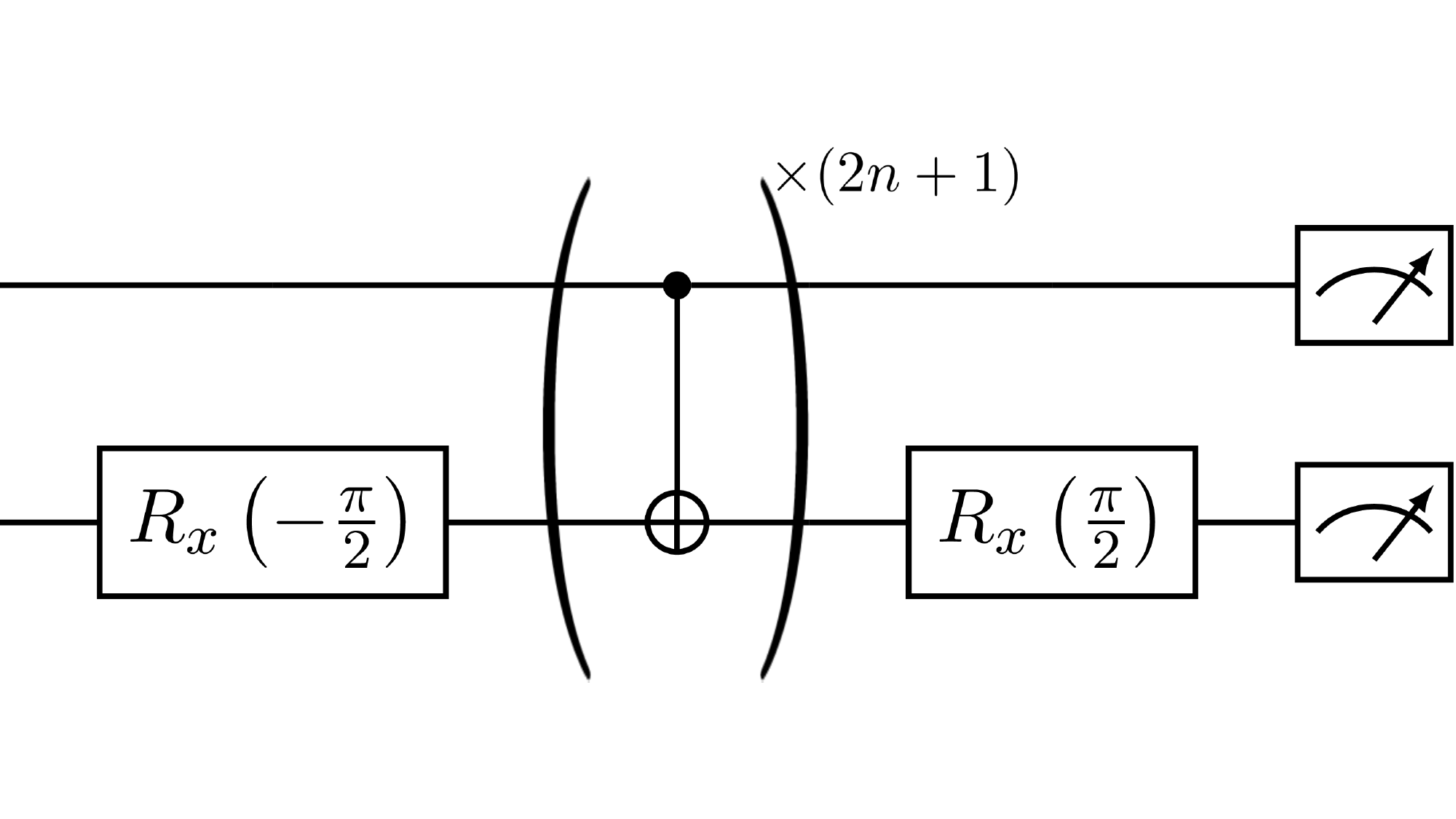}   & \includegraphics[height=2.8cm]{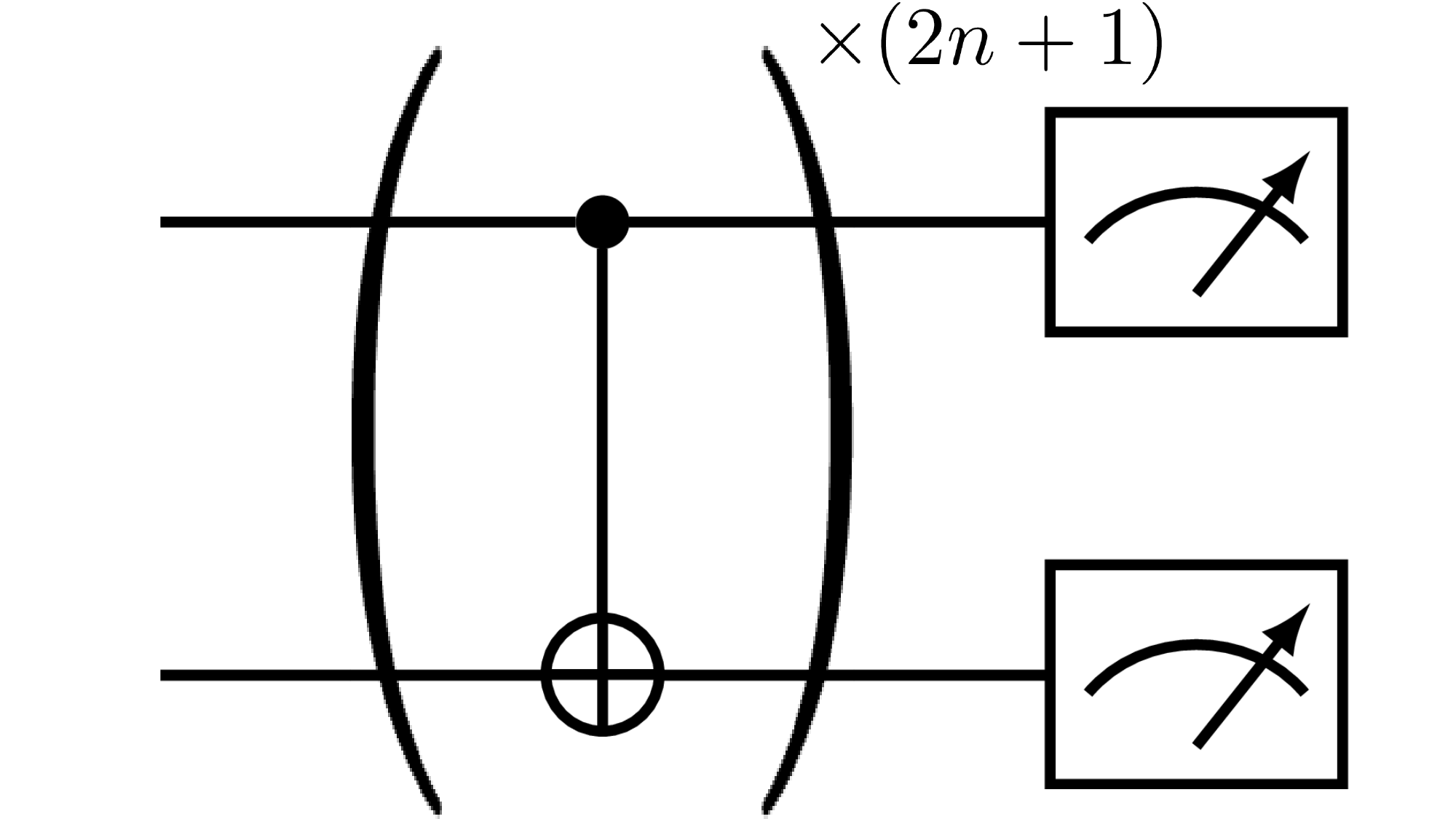} \\
\end{tabular}
    \caption{Circuits used for measuring  $(f_{\text{XX}}f_{\text{XI}})^nf_{\text{XX}}$, $(f_{\text{XX}}f_{\text{XI}})^nf_{\text{XI}}$ (top left), $(f_{\text{YX}}f_{\text{YI}})^nf_{\text{YX}}$, $(f_{\text{YX}}f_{\text{YI}})^nf_{\text{YI}}$ (top right),  $(f_{\text{IY}}f_{\text{ZY}})^nf_{\text{ZY}}$, $(f_{\text{IY}}f_{\text{ZY}})^nf_{\text{IY}}$ (bottom left) and $(f_{\text{IZ}}f_{\text{ZZ}})^nf_{\text{ZZ}}$, $(f_{\text{IZ}}f_{\text{ZZ}})^nf_{\text{IZ}}$ (bottom right) }
    \label{fig:cat2}
\end{figure*}

\begin{figure*}[h!]
    \centering
    \includegraphics[width=.7\linewidth,trim={0 2cm 0 0},clip]{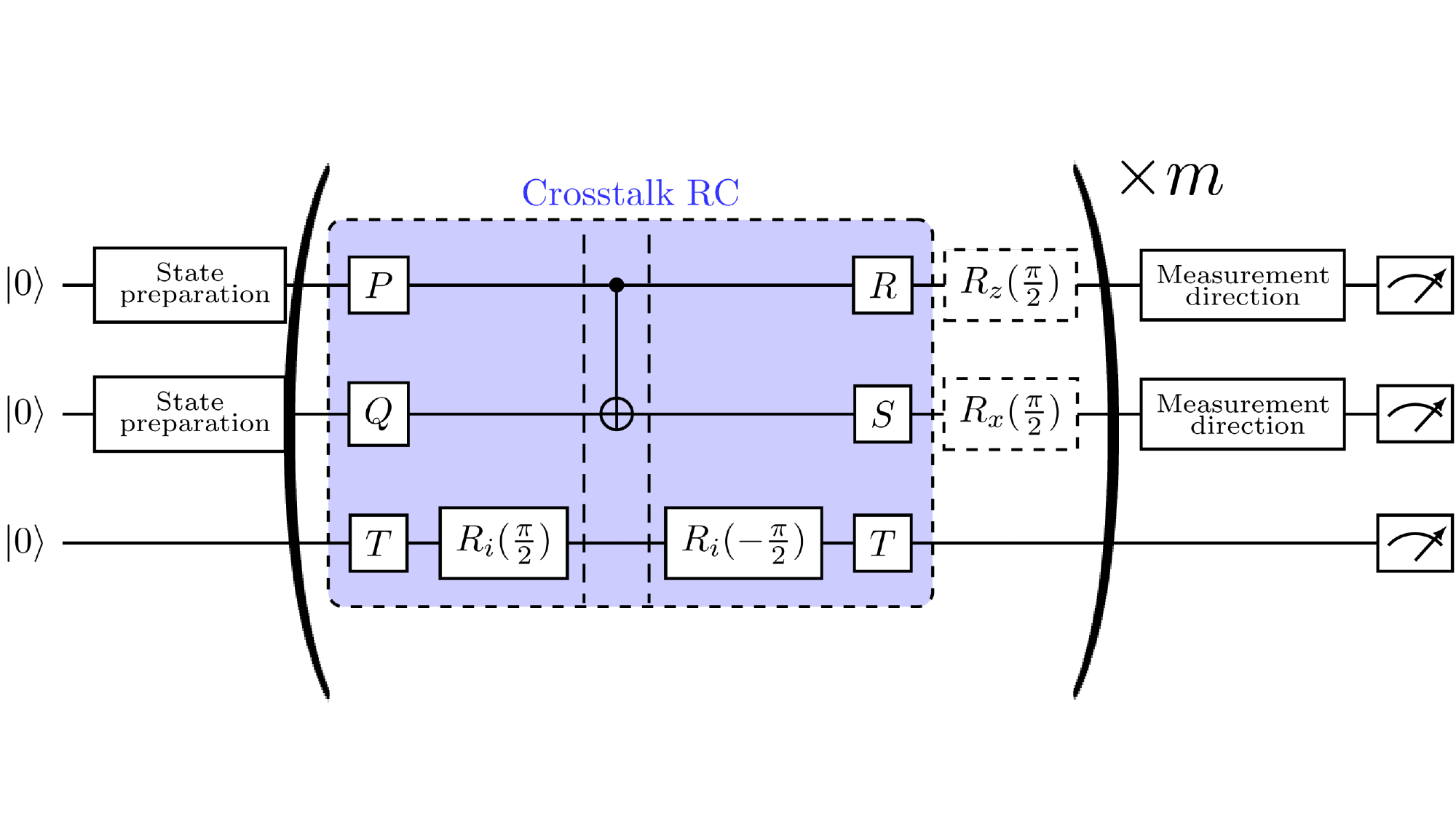}
    \caption{General structure of the full circuits used to characterize crosstalk noise. Preparation and measurement gates correspond to the nine circuits described in the Appendix~\ref{app:PNT}. Fully characterizing the crosstalk channel only requires a measurement on neighboring qubits but no additional circuits. Dotted $\pi/2$ rotation gate applied after RC are present only in four circuits out of nine (circuits of Fig.~\ref{fig:cat1}). The number of CNOT gates employed $m=2n$ is even for circuits of Fig.~\ref{fig:inv_fid},~\ref{fig:cat1} and $m=2n+1$ is odd for circuits of Fig.~\ref{fig:cat2}. }
    \label{fig:PNT_crosstalk}
\end{figure*}

\subsection{With crosstalk (cPNT)}
\label{app:cPNT}

We now explain how to extend the PNT protocol described above to the case of a CNOT with a single neighboring qubit, which is affected by crosstalk noise.

We recall (see Sec.~\ref{app:crosstalk}) that after performing cRC, the $3$-qubit noise channel under consideration has the following the PTM representation, a diagonal matrix of dimension 64:
\begin{equation}
       E_3^\textrm{gate}= \text{diag}(\textbf{F}^{\mathbb I},\textbf{F}^{\mathrm D},\textbf{F}^{\mathrm D},\textbf{F}^{\mathrm D})
        \label{eq:fid1}
\end{equation}
where
\begin{align}
\mathbf F^{\mathbb I}=\bigl[ f_{a{\mathbb I}}\bigr]_{a=0}^{15}\quad\text{and}\quad
\mathbf F^{\mathrm D}=\bigl[f_{a{\mathrm D}}\bigr]_{a=0}^{15}
\end{align}
are $16$-dimensional vectors of fidelities. The coefficients $f_{a\mathbb{I}}$ (resp. $f_{a\text{D}}$) denote the fidelities of the noise channel when no error (resp. a depolarizing error) occurs on the neighboring qubit. The depolarizing nature of the errors on the neighboring qubit simplifies the characterization of respective fidelities.

It is sufficient to prepare the neighboring qubit in the $\ket{0}$ state, apply cRC and measure it. Ignoring the measurement on the third qubit leads to the characterization of $\mathbf{F}^{\mathbb I}$ (exactly equivalent to the standard PNT without crosstalk) while taking it into account gives access to $\mathbf{F}^{\mathrm D}$. This leads to a generalization of the circuits presented in the last subsection to the ones presented in Fig.~\ref{fig:PNT_crosstalk}. This circuit structure illustrates the exact procedure of cRC+PNT used in the main text before the application of cNT.

We note that a form of PNT with crosstalk effects has been previously discussed in Ref.~\cite{vandenberg2023}. In that work, a dense brickwork layout of CNOT gates was used. Addressing crosstalk in such conditions required simplifying assumptions on the noise model (referred to as ``sparse model'') in order to avoid exponential complexity for characterization. By contrast, our method based on cRC does not require simplifying assumptions on the Pauli noise model, as by design it converts Pauli noise on the neighboring qubits to depolarizing noise. Moreover, our method can be straightforwardly extended to an arbitrary number of neighboring qubits, without the need to add new PNT circuits. Indeed, as explained above when switching from PNT to cPNT, the characterization can be accomplished by using the same set of circuits employed for noise characterization in the absence of crosstalk and simply extending the measurements to the neighboring qubits.

\section{Details about NISQ experiments and the noise used in their classical emulations}
\label{app:classical_simulations}

Here we discuss the procedure we used to run our protocol on IBMQ quantum computers and the computation cost of our protocol. We also discuss how our classical emulations incorporated the noise measured on the quantum computers.

\subsection{Selection of the best qubits at experiment time}
\label{app:selection}
Prior to the QC run, in a separate \texttt{qiskit-runtime} session, a complete cPNT tomography (see Appendix~\ref{app:PNT}) of the machine is conducted on each junction and in both directions (i.e. the CNOT gate is applied in both directions) in order to select the best three-qubit subset of the hardware. The cPNT protocol can be parallelized on distinct junctions as long as they don't share neighboring qubits. For \texttt{ibm\_hanoi} and \texttt{ibmq\_ehningen}, twelve separate parallel tomographies are necessary for a full noise profiling of the machine. For each of them, we run the nine different tomography circuits at four various depths $n=2^d$ with $d=0,1,2,3$ where $n$ is defined in Fig.~\ref{fig:inv_fid},~\ref{fig:cat1},~\ref{fig:cat2}. Each of these circuits is randomly compiled using $\tilde{N}_{\mathrm{RC}}=50$ circuits, and their output is sampled with $N_S=100$ shots. Analyzing these results, the optimal subset of three connected qubits is determined. Along with this tomography, we also require that each qubit has coherence times $T_1,T_2>2\times N_\text{CNOT}\times t_\text{CNOT}$ to ensure minimal decoherence effects, $N_\text{CNOT}$ being the number of CNOT gates in the circuit we wish to simulate and $t_\text{CNOT}$, the time duration for the application of a single CNOT gate. $t_\text{CNOT}$ and $T_1,T_2$ are provided by IBMQ's benchmarking data.

\subsection{Perfoming experiments in \texttt{Qiskit-runtime} sessions}
\label{app:runtime}
\par QC runs are organized into sessions, each providing exclusive access to the chosen quantum computer for a duration of 8 hours.

For the short simulation times,  $t\leq 2.6$, we use $1000$ circuits per time point for the total NT procedure. The circuit depth at these early simulation times is relatively low, allowing multiple time points to fit within a single \texttt{qiskit-runtime} session. With these simulation parameters, the quantum computation for all of the early time points fit in 4 sessions. For the final two simulation times however, $t = 2.8$ and $t = 3.0$, we employ $10000$ circuits per time point. In fact, we only present the QC data for the last two time points in this paper, cf.~Sec.~\ref{sec:exp_NISQ_protocol}.

\begin{figure}[h!]
    \centering
    \includegraphics[width=\linewidth]{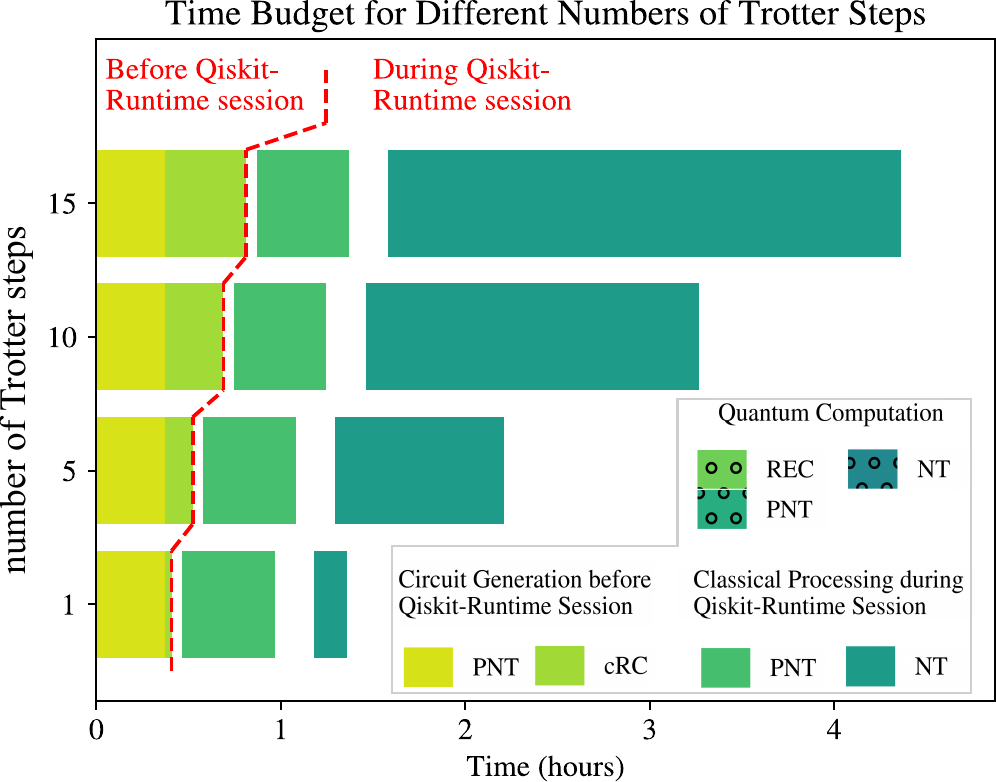}
    \caption{Wall-clock time spent on the various parts of the RC+NT+NEC protocol for different numbers of Trotter time steps (or equivalently simulation times, recall one Trotter time step is $\Delta t = 0.2$) in the case $1000$ circuits used for the RC+NT procedure. The most time-consuming part of the protocol for deep circuits is actually generating and uploading the NT circuits ("Classical processing"). Unfortunately, these cannot be generated in advance, as they depend on the noise of the device measured at the experiment time. The number of circuits required by different parts of the protocol is ran for the different method are detailed in Table~\ref{tab:overhead}. Note that the circuit generation of the PNT protocol before the \texttt{qiskit-runtime} session needs only to be performed once for all the Trotter time steps but is depicted on each of them for comparison.}
    \label{fig:time budget}
\end{figure}

The NT procedure, combined with the task of uploading the circuits, can be extremely time-consuming, as illustrated in Fig.~\ref{fig:time budget}. Thus, these two last time points require a total of 11 different \texttt{qiskit-runtime} sessions.

Before each of these \texttt{qiskit-runtime} sessions, a complete noise profile of the machine is constructed to determine the best subset of qubit as explained in Appendix~\ref{app:selection}. Consequently, qubits employed in different \texttt{qiskit-runtime} sessions may differ.

Additionally, the first hour of each \texttt{qiskit-runtime} session of the QC run is allocated to a more detailed cPNT (see Appendix~\ref{app:PNT}) procedure of the selected junctions. We employ $\tilde{N}_{\mathrm{RC}} = 200$ randomly compiled circuits at $n_d = 5$ different depths $n$ with $N_S = 100$ shots per circuit. The depths $n$ are chosen as follows: $n = 2^d$ with $d=0,1,2,3,4$.

\begin{figure}[h!]
    \centering
    \includegraphics[width=\linewidth]{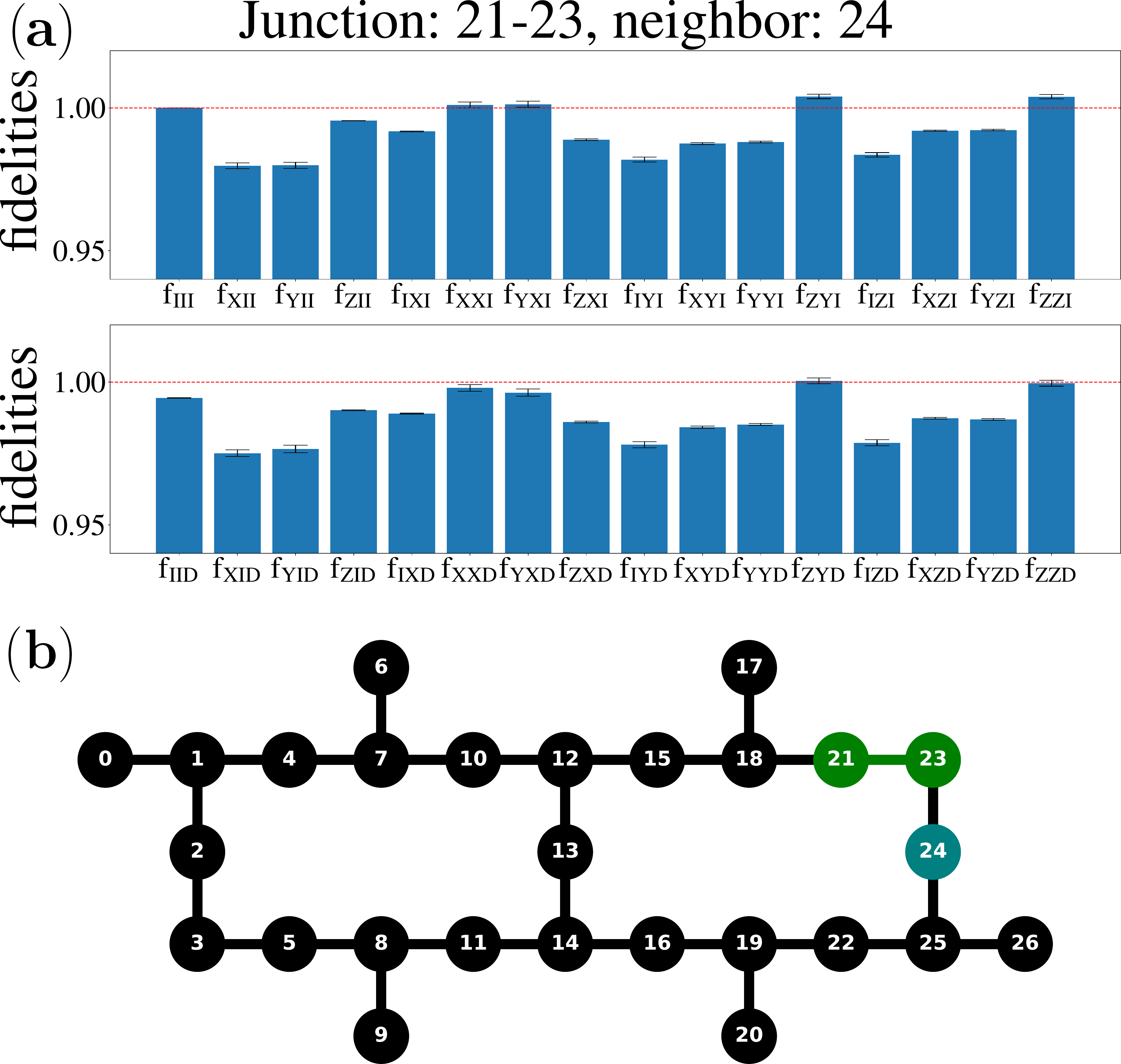}
    \caption{(a)  Example of the fidelities measured using cPNT at the beginning of a \texttt{qiskit-runtime} session, specifically focusing on the tomography of the CNOT gate applied to qubits 21 (control qubit), 23 (target qubit) (see the green qubits in panel (b)) and 24 (neigboring qubit, turquoise qubit in panel (b)) of the \texttt{ibmq\_ehningen} machine.  The first (resp. second) row displays the fidelities of $\mathbf{F}^{\mathbb I}$ (resp. $\mathbf{F}^{\mathrm D}$), cf.~Eq.~\eqref{eq:fid1}. The red line marks the value of 1, the upper limit of what can be considered meaningful fidelities. (b) Coupling map of \texttt{ibmq\_ehningen} machine. Links represent the physical junctions on which 2-qubit gates can be performed. The colored qubits are used for the simulation and black qubits are the idle qubits.}
    \label{fig:fidelites_crosstalk}
\end{figure}

An example of the fidelities derived from the cPNT for a single junction is shown in Fig.~\ref{fig:fidelites_crosstalk}(a). In the first row, we display the $16$ fidelities of $\mathbf{F}^{\mathbb I}$ involving only the two active qubits of the CNOT gate along each direction in Pauli space, see Eq.~\eqref{eq:fid1}. Within this row, the first bar $\text{f}_{\text{III}}$, is invariably set to one due to the trace-preserving condition and is not included in the count of the $15$ independent fidelities. The second row presents the set of $16$ fidelities of $\mathbf{F}^{\mathrm D}$, involving the neighboring qubit, see Eq.~\eqref{eq:fid1}. These fidelities are all computed using the circuits shown in Fig.~\ref{fig:cat2}. Error bars were obtained from 1,000 bootstrap resamples and account for both the shot noise and the finite RC sampling.

We  notice  that certain fidelities (in this example, $\text{f}_\text{XXI}$, $\text{f}_\text{YXI}$, $\text{f}_\text{ZYI}$ and $\text{f}_\text{ZZI}$) exceed the physical limit of $1$, sometimes beyond the error bar uncertainty. This phenomenon is highlighted in Fig.~\ref{fig:fidelites_crosstalk}(a), where a horizontal red line indicates the maximum fidelity level of 1: $\text{f}_\text{ZYI}$ and $\text{f}_\text{ZZI}$ clearly exceed it. Fidelities exceeding $1$ result in negative Pauli error probabilities $p_a$ for the corresponding Pauli noise channel. We have attempted to replicate this anomaly through noisy classical simulations that included coherent noise on two-qubit gates prior to RC, as well as noise on single-qubit gates, and even with noisy readout errors, yet were unsuccessful. We thus attribute this anomaly to the gauge freedom on the determination of Pauli noise from PNT, which can, in principle, be solved by formulating a self-consistent approach that includes SPAM errors, as discussed in Sec.~\ref{subsubsec:PNT}. In our classical emulations, we modify the Pauli error probabilities by setting the negative ones to 0. In our NISQ experiments, do not modify anything about the extracted noise channel, and use the extracted fidelities directly to calculate the required sampling in the NT protocol, cf.~App.~\ref{app:NT}.

We then generate NT circuits based on this noise profile, extracted via cPNT at the beginning of each \texttt{qiskit-runtime} session, and execute these circuits. Note that the generation of NT circuits (which ``resample" the inferred noise structure into the desired one) dominates the calculation time for deep circuits (large number of Trotter steps). While this is a completely classical part, it cannot be performed in advance, as it requires the knowledge of the noise on the hardware.

\subsection{Circuit overhead for our experiments}
\label{app:overhead}

\par Our protocol consists of several aspects: RC for converting the noise to the Pauli noise, PNT for characterizing the Pauli noise on the device, NT for tailoring the noise structure, REC for correcting the hardware errors at readout, and NEC for mitigating the simulation errors. Each of these stages requires running multiple circuits to achieve its goal. Table~\ref{tab:overhead} details the number of circuits needed for each step of our protocol.

\begin{table*}[ht!]
\begin{tabular}{|m{2cm}||c|m{3.5cm}|m{4cm}|m{4cm}|c|}
\hline
\centering method   & \centering cRC & \centering REC & \centering cPNT+cRC & \centering cNT+cRC & \centering NEC \arraybackslash \\
   \hline
   \hline
\begin{minipage}[c][1cm][c]{2cm}
        \centering
       number of circuits
    \end{minipage}   & $N_{\text{RC}}$ &\centering $2^{N_q} = 2^3 = 8$ & \begin{minipage}[c][1.2cm][c]{4cm}\centering $N_{\mathrm{PNT}} = 9 n_d \times 2 N_j \times \tilde{N}_{\mathrm{RC}}$
    \newline$n_d=5$, $\tilde{N}_{\mathrm{RC}}=200$, $N_j = N_q-1 = 2$ \end{minipage}  & \begin{minipage}[c][1cm][c]{4cm}\centering $N_\text{NT}=1000$ for $t \leq 2.6$ \newline \centering $N_\text{NT}=10000$ for $t \geq 2.8$ \end{minipage} &  
    \tikzmark{start1}\tikzmark{end1}
    \\\hline
    \begin{minipage}[c][1cm][c]{2cm}
        \centering number of shots per circuit\end{minipage} & \tikzmark{start2}\tikzmark{end2} & \centering $N_S=10^6$ &\centering $N_S=100$ & \centering $N_S=1000$ & \tikzmark{start3}\tikzmark{end3} \\
    \hline
    \centering statistical error & $\sigma_\text{RC} \propto \frac{1}{\sqrt{N_\text{RC}}}$ &
    \begin{minipage}[c][2.2cm][c]{3.5cm}
        \centering
        $\begin{aligned} \sigma_\text{REC} & \sim \sqrt{\frac{(1-\varepsilon_m)\varepsilon_m}{N_S}} \\ & \sim 10^{-4} \end{aligned} \newline \varepsilon_m$ is the measurement error rate
    \end{minipage}
    & \begin{minipage}[c][1cm][c]{4cm}
        \centering$\sigma_\text{PNT}$ is estimated\newline using jackknife resampling\newline$\sigma_\text{PNT} \sim 10^{-2}/10^{-3}$ \end{minipage}&
        \begin{minipage}[c][0cm][c]{4cm}
        \centering
        \vspace{0.9cm}
        $\sigma_\text{NT} \propto \frac{\gamma^{n_\text{CNOT}}}{\sqrt{N_\text{NT}}}$
        \end{minipage}

        & $\sigma_\text{mit}=\frac{\sigma_\text{NT}}{\mathcal{F}_\text{NEC}}$\\
    \hline
\end{tabular}

\caption{Table detailing the number of circuits and shots used for each method of the process as well as the scaling of the statistical error associated with.}
    \label{tab:overhead}

\begin{tikzpicture}[overlay, remember picture]
  \fill[pattern=north east lines, pattern color=black]
    ($(pic cs:start1)+(-1.cm,-0.5cm)$) rectangle
    ($(pic cs:end1)+(1.cm,0.65cm)$);

  \fill[pattern=north east lines, pattern color=black]
    ($(pic cs:start2)+(-0.99cm,-0.4cm)$) rectangle
    ($(pic cs:end2)+(1cm,0.55cm)$);
  \fill[pattern=north east lines, pattern color=black]
    ($(pic cs:start3)+(-1.cm,-0.4cm)$) rectangle
    ($(pic cs:end3)+(1.cm,0.55cm)$);
\end{tikzpicture}
\end{table*}

The first column of Table~\ref{tab:overhead} shows the statistical error, associated with RC whenever it is used (see the next columns). Indeed the accuracy of converting arbitrary noise to the Pauli noise is controlled by the number $N_\text{RC}$ of circuits used in RC. The second column corresponds to REC: one calibrates the readout matrix by preparing the qubit set in various states. The third column corresponds to the combined RC+PNT: one applies RC to the PNT circuits in order to characterize the Pauli noise obtained after the RC. The fourth column corresponds to the combined RC+NT. Indeed, NT must be combined with RC (cf.~Fig.~\ref{fig:workflow}) in order to convert the Pauli noise to the target noise. Note that we used an increased number of circuits for the last two time points in our simulations. Finally, the error mitigation by NEC, as shown in the last column, requires having to estimate the fidelity $\mathcal{F}_\text{NEC}$. This circuit was not run on a QC, but was emulated classically, as it is a Clifford circuit and the target Pauli noise of NT is known. Division of the result by $\mathcal{F}_\text{NEC}$ amplifies the statistical error of the estimated observable.

One can see from the table that our protocol requires a large number of distinct NT circuits. Furthermore, these circuits can only be generated once the noise characterization has been performed. Figure~\ref{fig:time budget} shows that for deep circuits this classical processing takes up the majority of time of the quantum \texttt{qiskit-runtime} session, wasting the time of access to the quantum computer. For instance, the generation of each of the deepest circuit used in our experiment takes on average 10 seconds. We believe that optimization of circuit handling may speed up this part significantly. We, however, did not investigate this direction in practice.

\subsection{Pauli noise used in classical emulation}
\label{app:noise_in_classical_emulations}

In order to maximize the correspondence between classical emulations and NISQ experiments, he Pauli noise channels used for the classical emulations stem from the output of the PNT procedure performed in NISQ experiments just before executing the NT protocol. In particular, if a simulation of a specific time point was distributed over several distinct \texttt{qiskit-runtime} sessions and thus involved running circuits under different noise conditions, we classically emulate the very same numbers of circuits run under the respective noise conditions.
As discussed in App.~\ref{app:runtime}, cPNT tomography may return fidelities exceeding $1$, yielding nonphysical (negative) Pauli error probabilities. Since classical emulation cannot reproduce these negative probabilities, we set such probabilities $p_a$, cf.~Eq.~\eqref{eq:pauli_noise}, to zero and renormalize the other probabilities to preserve the normalization of the channel, $\sum_a p_a = 1$. Note that this is different from setting the fidelities $f_a$ exceeding 1 to 1. Therefore, the NT sampling in our NISQ runs (where we directly use the fidelities extracted by the cPNT) and classical emulations differs slightly. At the same time, the RC sampling in classical emulations is identical to that in the \texttt{qiskit-runtime} sessions on real QC.

Note also, that our classical emulations cannot replicate residual coherent noise or violations of assumptions (i) and (ii) discussed in Sec.~\ref{sec:raw_noise}, as we do not possess the means to characterize them. Therefore, we attribute the major discrepancies between the NISQ results and classical emulations to these types of noise.

\section{Results for ibmq\_ehningen}
\label{app:ehningen_results}

\begin{figure*}[t]
  \centering
  \begin{minipage}[t]{0.48\textwidth}
    \vspace{0pt}
    \centering
    \includegraphics[width=\linewidth]{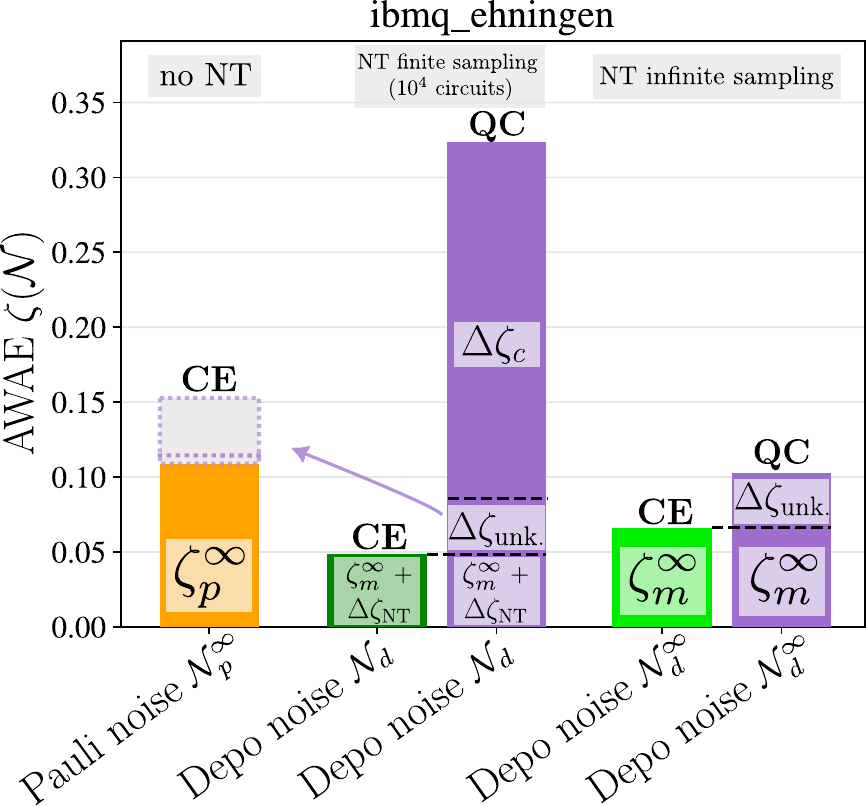}
    \caption{Comparison of the average weighted absolute error $\zeta$ for the last two time points in classical emulations (CE) and in actual quantum computer runs (QC). Purple bars show the QC results obtained on \texttt{ibmq\_ehningen}, and the green and the orange bars show the classical emulation results from Fig.~\ref{fig:error_class_sim}. The purple bar in the center shows $\zeta$ obtained in the quantum computer runs with RC+NT+NEC protocol using $N_{\text{NT}} = 10^4$ sampling circuits. Comparing it to the dark-green bar in the center one identifies the contributions of the residual coherent and unknown noise channels in the QC runs. The dashed purple blocks on top of the orange bar show the estimated contribution of these noise sources in the RC+NEC protocol, see Sec.~\ref{sec:accuracy_non-improvement_NT+NEC} for details. The purple bar on the right shows an estimate for AWAE in the limit of $N_{\text{NT}} = \infty$, which eliminates the residual coherent noise completely; the extrapolation procedure producing this estimate is described in Sec.~\ref{sec:extrapolation}.}
    \label{fig:error ehningen}
  \end{minipage}\hfill
  \begin{minipage}[t]{0.48\textwidth}
  \vspace{0pt}
    \centering
    \includegraphics[width=\linewidth]{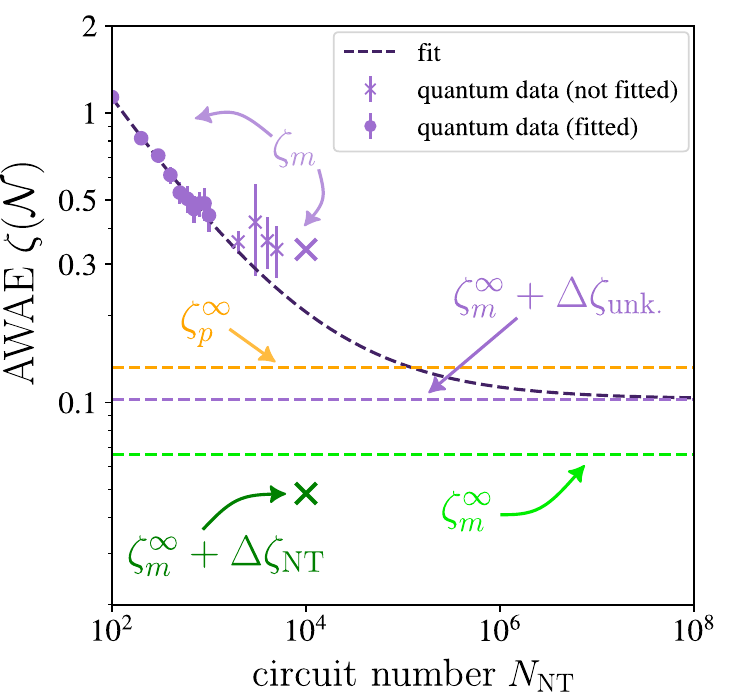}
    \caption{The AWAE $\zeta$ of the \texttt{ibmq\_ehningen} quantum computer runs extrapolated to the infinite-sampling limit. The purple dots and crosses show the $\zeta$ levels for various number of sampling circuits estimated through bootstrapping. The dashed black curve shows the result of fitting the data points with Eq.~\eqref{eq:AWAE_fit_formula}. The horizontal purple dashed line shows the extrapolated $\zeta$ level for infinite sampling, i.e., $b$ from Eq.~\eqref{eq:AWAE_fit_formula}. Other $\zeta$ levels from Fig.~\ref{fig:error ehningen} are shown for reference.}
    \label{fig:scaling ehningen}
  \end{minipage}
\end{figure*}

The results for \texttt{ibmq\_ehningen} are presented in Fig.~\ref{fig:error ehningen}. We perform the same analysis as in the main text for \texttt{ibm\_hanoi}, which allows us to identify the contribution of the different noise types to the AWAE. While the quantum results display a trend similar to the results of \texttt{ibm\_hanoi}, the magnitude of the correction $\Delta\zeta_c$ is significantly larger on this quantum computer.

In Fig.~\ref{fig:scaling ehningen}, we also present the extrapolation of the AWAE $\zeta$ as a function of the number of circuits for $\texttt{ibmq\_ehningen}$, from which we extract both $\Delta\zeta_{\textbf{unk.}}$ and $\Delta\zeta_c$. Surprisingly, increasing the number of circuits from 2000 to 10000 does not lead to a reduction in the AWAE. Although this might be attributed to statistical fluctuations, given the significance of the error bars, it may also reflect the instability of the current NISQ devices in terms of experimental noise, as already observed in the literature for IBMQ devices~\cite{woitzik2024,Dasgupta2024,Filippov2024}.

\end{document}